\documentclass[seceq]{ptptex}

\usepackage{amsmath}
\usepackage{amssymb}
\usepackage[dvipdfm]{graphicx}
\usepackage{subfigure}

\preprintnumber{IU-TH-7}

\title{
Formulation of Complex Action Theory
}

\author{%
Keiichi \textsc{Nagao}\footnote{E-mail: nagao@mx.ibaraki.ac.jp, nagao@nbi.dk} 
and Holger Bech \textsc{Nielsen}\footnote{E-mail: hbech@nbi.dk}
}

\inst{
$^{*)}$Faculty of Education, Ibaraki University, Mito 310-8512 
Japan \\
{\it and }\\
$^{*),**)}$Niels Bohr Institute, University of Copenhagen, 
17 Blegdamsvej Copenhagen $\phi$, Denmark
}

\abst{

We formulate a complex action theory which includes operators of coordinate and momentum 
$\hat{q}$ and $\hat{p}$ being replaced with 
non-hermitian operators $\hat{q}_{new}$ and 
$\hat{p}_{new}$, and their eigenstates ${}_m \langle_{new}~ q |$ and ${}_m \langle_{new}~ p |$ with 
complex eigenvalues $q$ and $p$. 
Introducing a philosophy of 
keeping the analyticity in 
path integration variables, 
we define a modified set of complex conjugate, real and imaginary parts, hermitian conjugates and bras, 
and explicitly construct $\hat{q}_{new}$, $\hat{p}_{new}$, ${}_m \langle_{new}~ q |$ and ${}_m \langle_{new}~ p |$ 
by formally squeezing coherent states. 
We also pose a theorem on the relation between functions on the phase space 
and the corresponding operators. 
Only in our formalism can we describe a complex action theory or 
a real action theory with complex saddle points in the tunneling effect 
etc. in terms of bras and kets in the functional integral.  
Furthermore, in a system with a non-hermitian diagonalizable bounded Hamiltonian, 
we show that the mechanism to obtain 
a hermitian Hamiltonian after a long time development proposed in our letter~\cite{Nagao:2010xu} 
works also in the complex coordinate formalism. 
If the hermitian Hamiltonian is given in a local form, a conserved probability current density 
can be constructed with two kinds of wave functions. 
}

\begin{document}

\maketitle

\section{Introduction} \label{Intro}


Feynman path integral (FPI) is one of the essential routes to formulate quantum 
theories. 
In quantum theory with an action $S$ the integrand in FPI has the form 
of $\exp(\frac{i}{\hbar}S)$, 
where $i$ is the imaginary unit. 
Usually S is real, and it is thought to be more fundamental than the 
integrand. 
However, if we assume that the integrand is more fundamental
than the action in quantum theory, then it is naturally thought that
since the integrand is complex, the action could be also complex. 
Based on this assumption and other related works 
in some backward causation developments inspired by general relativity\cite{ownctl} 
and the non-locality explanation 
of fine-tuning problems \cite{nonlocal}, the complex action theory (CAT) 
has been studied intensively by one of the authors (H.B.N) and Ninomiya\cite{Bled2006,own}. 
Compared to the usual real action theory (RAT), the imaginary part of the action
is thought to give some falsifiable predictions. 
Indeed, many interesting suggestions have been made for Higgs mass\cite{Nielsen:2007mj}, 
quantum mechanical philosophy\cite{newer}, 
some fine-tuning problems\cite{Nielsen2010qq,degenerate}, 
black holes\cite{Nielsen2009hq}, 
De Broglie-Bohm particle and a cut-off in loop diagrams\cite{Bled2010B}.

In refs.\cite{Bled2006,own,Nielsen:2007mj,newer,Nielsen2010qq,degenerate,Nielsen2009hq,Bled2010B} 
they studied a future-included version, that is to say, 
the theory including not only a past time but also a future time 
as an integration interval of time. 
In contrast to them, in ref.\cite{Nagao:2010xu}  we 
have studied in a future-not-included version 
the time-development of some state by a non-hermitian diagonalizable bounded Hamiltonian $H$. 
As for non-hermitian Hamiltonians, the formalism based on 
the PT symmetry has been intensively studied in both theoretical aspects\cite{PTsym_Hamiltonians,Geyer} 
and experimental ones\cite{Experiments}. 
The eigenvalues are real, 
and such a PT symmetry has been considered also in a different context\cite{Erdem:2006zh}. 
On the other hand, the Hamiltonian we have studied in ref.\cite{Nagao:2010xu} is 
generically non-hermitian, so its eigenvalues are complex in general. 
In addition, since the eigenstates are not orthogonal, 
a transition that should not be possible could be measured. 
From these properties it does not look a physically reasonable theory, 
but we have proposed a framework to obtain a hermitian Hamiltonian 
effectively based on the speculation in ref.\cite{originsym}. 
The framework is composed of two steps: 
As the first step we have defined 
a physically reasonable inner product $I_Q$ such that 
the eigenstates of the Hamiltonian get orthogonal with regard to it. 
Then it gives us a true probability for a transition from some state to another. 
With regard to $I_Q$ the Hamiltonian is normal and we have defined a hermiticity 
with regard to it, $Q$-hermiticity. 
A similar inner product has been studied also in ref.\cite{Geyer}. 
As the second step we have presented a mechanism of 
suppressing the effect of the anti-hermitian part of the Hamiltonian $H$ 
after a long time development. 
For the states with high imaginary part of eigenvalues of $H$, 
the factor $\exp\left(-\frac{i}{\hbar} H(t - t_0) \right)$ grows exponentially with $t$. 
After a long time the states with the highest imaginary part of eigenvalues of $H$ 
get more favored to result than others. 
Thus, the effect of the imaginary part, namely the anti-$Q$-hermitian part of $H$, 
gets attenuated except for an unimportant constant. 
Utilizing this effect to normalize the state, 
we have obtained a $Q$-hermitian Hamiltonian effectively. 
We have also constructed a conserved probability current density with two kinds of wave functions 
under the assumption that the $Q$-hermitian Hamiltonian is given in a local form. 
Furthermore we have pointed out a possible misestimation of a past state 
by extrapolating back in time with the hermitian Hamiltonian.

In addition, as other works related to complex saddle point paths, in refs.~\cite{Garcia:1996np}\cite{Guralnik:2007rx} 
the complete set of solutions of the differential equations 
following from the Schwinger action principle has been obtained by generalizing 
the path integral to include sums over various inequivalent contours of integration in 
the complex plane. 
In ref.~\cite{Pehlevan:2007eq} complex Langevin equations have been studied. 
In refs.~\cite{Ferrante:2008yq}\cite{Ferrante:2009gk} a method to examine the complexified solution set  has been investigated.

The CAT has been studied intensively as mentioned above. 
But there still remain many things to be investigated. 
Once we allow the action to be complex, various quantities known 
in the RAT can drastically change. 
For example, in the CAT a coordinate $q$ and a momentum $p$ obtained at saddle points 
can be complex, so we could encounter various exotic situations. 
Also, we note that there would be some ways to extend the action to complex: 
whether mass and other coupling parameters are complex or not,  
whether $q$ and $p$ are generically complex or not, 
and also whether we include $q^*$ in the action or not, etc. 
In this manuscript we formulate the CAT 
such that mass and other coefficients are generically complex, 
while dynamical variables such as $q$ or $p$ are fundamentally real but can be complex at saddle points. 
Already in the RAT, we find classical paths (corresponding to 
saddle points in the functional integral) along which the dummy variables $q(t)$ say take on 
complex values. 
As long as we just have an analytic form for $S(path)$ 
as a function of $q(t)$ (for all $t$), we are naturally allowed to deform a path of integration 
without changing the functional integral. 
Usually Dirac derives the functional integral by inserting 
the completeness relation $\int_{-\infty}^\infty | q \rangle \langle q | dq = 1$ 
into the matrix element $\langle q(t_f) | e^{-iHt} | q(t_i) \rangle$, but 
after the deformation when $q(t)$ should be complex, say in such cases as 
the tunneling effect or the WKB approximation etc., 
the symbols $| q(t) \rangle$ and $\langle q(t) |$ have not been used. 
At first there is a very good reason for being reluctant to write down $| q(t) \rangle$ and $| p(t) \rangle$ 
for complex $q(t)$ and $p(t)$: 
the operators of coordinate and momentum $\hat{q}$ and $\hat{p}$ are hermitian 
and thus have only real eigenvalues. 
To get complex eigenvalues for them allowed 
we replace them by non-hermitian operators $\hat{q}_{new}$ and $\hat{p}_{new}$ 
approaching $\hat{q}$ and $\hat{p}$ only in certain limits of parameters present in the definitions of 
$\hat{q}_{new}$ and $\hat{p}_{new}$ (See section~\ref{construction}).  
The main purpose of this paper is to allow a formulation of quantum theory in terms of 
$\hat{q}_{new}$, $\hat{p}_{new}$ and their eigenstates ${}_m \langle_{new}~ q |$ and ${}_m \langle_{new}~ p |$ 
with {\em complex} eigenvalues $q$ and $p$, i.e. with eigenvalue 
equations ${}_m \langle_{new}~ q | \hat{q}_{new} = {}_m \langle_{new}~ q | q$ and 
${}_m \langle_{new}~ p | \hat{p}_{new} = {}_m \langle_{new}~ p | p$. 
Here ${}_m\langle_{new}~  q | $ and ${}_m\langle_{new}~ p | $ are modified bras of 
$| q \rangle_{new}$ and $| p \rangle_{new}$, which are define to keep the analyticity in $q$ and $p$ 
respectively, so the two relations are equivalent to 
$\hat{q}_{new}^\dag | q \rangle_{new} = q | q \rangle_{new}$ and 
$\hat{p}_{new}^\dag | p \rangle_{new} = p | p \rangle_{new}$.   
Unless we replace $\hat{q}$ and $\hat{p}$ by 
the slightly modified operators $\hat{q}_{new}$ and $\hat{p}_{new}$, 
we cannot have complex eigenvalues. 
Thus it is only with the replacement 
that we can be allowed to write down the eigenstates ${}_m\langle_{new}~  q |$ and ${}_m\langle_{new}~  p |$ 
for complex eigenvalues $q$ and $p$.

In FPI we would like to be allowed to deform contours of integration over $q$ or $p$ 
to complex contours passing saddle points keeping the endpoints {$\mp \infty$} on the real axis. 
We assume that the asymptotic values of $q$ and $p$ are real.\footnote{In principle our CAT is a quantum theory from the start having only real $q$ values. 
Thus any ``wave packet" around a complex center $q' \in {\bold C}$ would in principle be a somewhat 
complicated quantum state with real $q$. It is only for convenience in studying 
the CAT that we suggest to play formally with complex $q$-eigenstates. Our CAT is already 
quantized in our basic formulation; so we do not have to quantize again, and it is certainly not needed to 
do so using complex $q$-states (after all only real ones exist fundamentally). } 
Indeed we shall make such contour deformation possible by insisting on introducing the philosophy 
of keeping the analyticity in dynamical variables of FPI. 
To realize this philosophy we define a modified set of complex conjugate, 
real and imaginary parts, hermitian conjugates and bras. 
Also, we study the delta function of $q$ say and show that even for a complex parameter $q$ 
it behaves as the usual delta function if $q$ satisfies the condition 
$\left( \text{Re}(q) \right)^2 > \left( \text{Im}(q) \right)^2$. 
Based on this philosophy we can deform a path in the complex parameter plane in FPI. 
If we choose so, we can settle the path down to the real axis 
in the complex plane and the extension is in this way formal. 
Thus a CAT can be interpreted -- at least quantum mechanically -- in terms of fundamentally real 
$q$ and $p$, and in principle 
the CAT is falsifiable by giving predictions that can be compared to 
experiments with of course real dynamical variables. 
We explicitly show one example of constructing the non-hermitian operators $\hat{q}_{new}$ and $\hat{p}_{new}$, 
their eigenstates ${}_m\langle_{new}~  q |$ and ${}_m\langle_{new}~  p |$ with complex $q$ and $p$ 
by formally squeezing coherent states of harmonic oscillators, 
and see that they satisfy the relations 
$\hat{p}_{new}^\dag | q \rangle_{new} = i \hbar \frac{\partial   | q \rangle_{new} }{\partial q}$ and 
$\hat{q}_{new}^\dag | p \rangle_{new} = \frac{\hbar}{i} \frac{\partial  | p \rangle_{new} }{\partial p}$ 
by insisting on $[\hat{q}_{new},  \hat{p}_{new} ] = i \hbar$. 

Furthermore, to make clear the relation between functions on the phase space 
describing some classical variables 
and the corresponding operators under quantization, 
providing the notions of ``$\epsilon$-analytical" functions, 
and ``expandable" and  ``$\epsilon$-expandable" operators, we pose a theorem, 
which claims that if and only if some operator corresponding to an $\epsilon$-expandable 
function on the phase space, its matrix element in $q$-representation is an $\epsilon$-analytical function. 
In addition, as an application of the complex coordinate formalism 
we attempt to extend the mechanism proposed in ref.\cite{Nagao:2010xu} 
to the complex coordinate formalism. 
We study a system defined by a diagonalizable non-hermitian bounded Hamiltonian,  
and show that the mechanism to obtain a hermitian Hamiltonian effectively 
after a long time development works also in the complex coordinate formalism. 
We see that if the hermitian Hamiltonian is given in a local form, a conserved probability current density 
can be constructed with two kinds of wave functions.

This paper is organized as follows. 
In section 2 we explain our proposal of replacing hermitian operators $\hat{q}$, $\hat{p}$ and their eigenstates 
$\langle q |$ and $\langle p |$ 
with $\hat{q}_{new}$, $\hat{p}_{new}$, ${}_m\langle_{new}~  q |$ and ${}_m\langle_{new}~  p |$. 
Introducing a philosophy of keeping the analyticity in dynamical variables of FPI 
we define a modified set of complex conjugate, real and imaginary parts, hermitian conjugates and bras. 
We also study the delta function of a complex parameter. 
In section 3 we explicitly construct 
$\hat{q}_{new}^\dag$ and $\hat{p}_{new}^\dag$, and their eigenstates $| q \rangle_{new}$ and $| p \rangle_{new}$ 
with complex eigenvalues $q$ and $p$ by formally utilizing coherent states of harmonic oscillators. 
In section 4, as an application of the complex coordinate formalism 
we extend the mechanism proposed in ref.\cite{Nagao:2010xu} 
to the complex coordinate formalism. 
Section 5 is devoted to summary and outlook. 
In appendix~\ref{app_cs} we briefly review a coherent state. 
In appendix~\ref{explicitcalc} we explicitly study various properties of 
$\hat{q}_{new}^\dag$, $\hat{p}_{new}^\dag$, $| q \rangle_{new}$ and $| p \rangle_{new}$. 
In appendix~\ref{theorem} 
we pose a theorem on the relation between functions on the phase space 
and the corresponding operators.

\section{Our proposal and new devices}  \label{fundamental}

In this section we first present our proposal 
and a philosophy of keeping the analyticity in dynamical variables of FPI. 
Next we introduce new devices to realize the philosophy, 
i.e. a modified set of complex conjugate, real and imaginary parts, 
hermitian conjugates and bras. 
We also study the delta function of a complex parameter.

\subsection{Our proposal and a philosophy of keeping the analyticity in dynamical variables}

We formulate the CAT 
such that mass and other coupling parameters are generically complex, 
while $q$ and $p$ are fundamentally real but can be complex at saddle points. 
As we have explained in section~\ref{Intro}, we encounter complex $q$ or $p$ 
not only in the CAT but also in the RAT, while $\hat{q}$ and $\hat{p}$ 
are hermitian and thus have only real eigenvalues. 
To get complex eigenvalues for them allowed 
we propose replacing hermitian operators $\hat{q}$ and $\hat{p}$, 
and their eigenstates $\langle q |$ and $\langle p |$ with non-hermitian operators 
$\hat{q}_{new}$, $\hat{p}_{new}$ and their eigenstates ${}_m\langle_{new}~  q |$ and ${}_m\langle_{new}~  p |$, 
which satisfy the following relations for complex $q$ and $p$, 
\begin{eqnarray}
&&{}_m \langle_{new}~ q | \hat{q}_{new} = {}_m \langle_{new}~ q | q , \label{qhatqket=qqket_newrev} \\
&&{}_m \langle_{new}~ p | \hat{p}_{new} = {}_m \langle_{new}~ p | p , \label{phatpket=ppket_newrev} \\ 
&&[\hat{q}_{new}, \hat{p}_{new} ] = i \hbar, \label{commutator_q_p_new} 
\end{eqnarray}
where ${}_m\langle_{new}~  q | $ and ${}_m\langle_{new}~ p | $ are modified bras of 
$| q \rangle_{new}$ and $| p \rangle_{new}$. We define modified bras later. 
Eqs.(\ref{qhatqket=qqket_newrev})(\ref{phatpket=ppket_newrev}) are equivalent to 
\begin{eqnarray}
&&\hat{q}_{new}^\dag  | q \rangle_{new} =q | q \rangle_{new} , \label{qhatqket=qqket_new} \\
&&\hat{p}_{new}^\dag  | p \rangle_{new} =p | p \rangle_{new} . \label{phatpket=ppket_new} 
\end{eqnarray}
In section \ref{construction} we explicitly construct them 
by formally utilizing coherent states of harmonic oscillators so that we can have complex eigenvalues.

In addition, we introduce a philosophy of keeping the analyticity in dynamical variables 
such as $q$ and $p$ of FPI. 
Then we can deform an integration path in the complex plane of $q$ or $p$, and 
$\hat{q}_{new}$ and $\hat{p}_{new}$ have eigenvalues on any deformed path. 
To realize this philosophy we shall define 
a modified set of complex conjugate,  real and imaginary parts, hermitian conjugates and bras 
in the following subsections.


\subsection{Modified complex conjugate $*_{ \{ \} }$ }

A usual complex conjugate is defined for a function of $n$-parameters 
$f( \{a_i \}_{i=1, \ldots, n} )$ as follows, 
\begin{equation}
f(\{a_i \}_{i=1, \ldots, n} )^{*} = f^*( \{a_i^* \}_{i=1, \ldots, n} ) ,
\end{equation}
where on the right-hand side 
$*$ on $f$ acts on the coefficients included in $f$. 
We introduce a modified complex conjugate as follows, 
\begin{equation}
f(\{a_i \}_{i=1, \ldots, n} )^{*_{\{a_i | i \in A \}} } = f^*( \{a_i \}_{i \in A}  ,  \{a_i ^*\}_{i \not\in A} ) , 
\end{equation}
where $A$ denotes the set of indices attached with the parameters, in which we keep the analyticity. 
By this newly defined complex conjugate we do not take complex conjugate of the parameters 
which we denote as subscripts of $*$. 
For example, if we are given $f(q,p)=a q^2 + b p^2$, then the following relations hold, 
\begin{eqnarray}
&&f(q,p)^{*_q} = f^*(q, p^*) = a^* q^2 + b^* (p^*)^2 , \\
&&f(q,p)^{*_{q,p}} = f^*(q, p) = a^* q^2 + b^* p^2 ,
\end{eqnarray}
where in the first and second relations the analyticity is kept in $q$,  
and both $q$ and $p$, respectively. 
For simplicity we express the modified complex conjugate as $*_{ \{  \} }$. 
%

\subsection{Modified real and imaginary parts $\text{Re}_{\{ \}}$, $\text{Im}_{\{ \}}$ }

We define the modified real and modified imaginary parts by using $*_{ \{  \} }$.
Indeed we can decompose some complex function $r$ as 
\begin{equation}
r= \text{Re}_{\{ \}} r + i \text{Im}_{\{ \}} r ,
\end{equation}
where $\text{Re}_{\{ \}} r$ and $\text{Im}_{\{ \}} r$ are defined by 
\begin{eqnarray}
&&\text{Re}_{\{ \}} r = \frac{ r + r^{*_{\{ \}}}   }{2} , \\
&&\text{Im}_{\{ \}} r = \frac{ r - r^{*_{\{ \}}}   }{2i} . 
\end{eqnarray}
They are the ``$\{ \}$-real" and ``$\{ \}$-imaginary" parts of $r$, respectively. 
For example, if we are given $r=\frac{1}{2}k q^2$, then we have 
\begin{eqnarray}
&&\text{Re}_{q} \left( \frac{1}{2} k q^2 \right) = \frac{1}{2} \text{Re} ( k ) q^2 , \\
&&\text{Im}_{q} \left( \frac{1}{2}k q^2 \right)  =  \frac{1}{2} \text{Im} ( k ) q^2 . 
\end{eqnarray}
Especially, if some complex number $r$ satisfies the following relation,
\begin{equation}
r^{*_{\{ \}}} =r , \label{def_ext_reality}
\end{equation}
we say $r$ is $\{ \}$-real, while if $r$ obeys 
\begin{equation}
r^{*_{\{ \}}} =-r , \label{def_ext_reality2}
\end{equation}
we call $r$ purely $\{ \}$-imaginary. 


\subsection{Modified bra ${}_m \langle ~|$, ${}_{ \{ \} } \langle ~|$ and modified hermitian conjugate $\dag_{ \{ \} }$}
\label{moreon}

For some state $| \lambda \rangle$ with some complex parameter $\lambda$, 
we define a modified bra ${}_m\langle \lambda |$ by 
\begin{equation} 
{}_m\langle \lambda | = 
\langle \lambda^* | . \label{modified_bra_anti-linear}
\end{equation}
This is an analytically extended bra with regard to the parameter $\lambda$. 
In the special case of $\lambda$ being real this becomes a usual bra. 
We introduce a little bit generalized modified bra, 
${}_{\{\}}\langle ~|$, where $\{ \}$ is a symbolical expression of a set of parameters 
in which we keep the analyticity. 
We show two examples, 
\begin{eqnarray}
&&{}_{u,v} \langle u | = {}_u \langle u | = {}_m\langle u | , \label{mod_bra2} \\
&&{}_u \langle v | = \langle v | , 
\end{eqnarray}
where $u$ and $v$ are some complex parameters.

We also introduce a modified hermitian conjugate $\dag_{ \{ \} }$ of a ket. 
This is an analytically extended hermitian conjugate with regard to the set of  parameters $\{ \} $. 
Then we can express the modified hermitian conjugate $\dag_{ \{ \} }$ of a ket as 
\begin{equation}
( |  ~\rangle )^{\dag_{\{ \} }} = {}_{\{ \}}\langle  ~| , 
\end{equation}
and we have the following relation 
\begin{equation}
( {}_{\{ \}}\langle  ~| )^{\dag_{\{ \} }} =  |  ~\rangle . 
\end{equation}
We show two examples,  
\begin{eqnarray}
&&( | u \rangle )^{\dag_{v }}  = ( | u \rangle )^\dag  = \langle u | , \\
&&( | u \rangle )^{\dag_{u, v }}  =( | u \rangle )^{\dag_{u}}  = {}_m \langle u |  = {}_{u,v}\langle u | .
\end{eqnarray}

Next we consider a hermitian conjugate $\dag_{\{ \} }$ of operators. 
In the RAT a hermitian conjugate of some operator $A$, $A^{\dag}$, is defined by the relation 
\begin{equation}
\langle u | A | v \rangle^{*} = \langle v | A^{\dag} | u \rangle . \label{hermitian_conjugate_RAT}
\end{equation}
In the CAT we extend it as 
\begin{equation}
{}_{\{\}}\langle u | A | v \rangle^{*_{ \{ \} }} = 
{}_{ \{ \} } \langle v | A^\dag | u \rangle , \label{hermitian_conjugate_5}
\end{equation}
and we have the following relation, 
\begin{equation}
A^{\dag_{\{\}} } = A^\dag . \label{herm_conj_op2}
\end{equation}
%

\subsection{The delta function}

The delta function is one of the essential tools in a theory which has 
orthonormal basis with continuous parameters, and the parameters are usually real in the RAT. 
In the CAT parameters are complex in general, 
so we attempt to extend the delta function  
to complex parameters.

The delta function is defined as a distribution by the relation 
\begin{equation}
\int f(q) \delta(q-a) dq = f(a),
\end{equation}
where $f(q)$ is a test function. 
By expanding $f(q)$ in Fourier series the delta function is represented as 
\begin{equation}
\delta(q) = \frac{1}{2\pi} \int_{-\infty}^\infty  e^{ikq} dk . \label{delta-func}
\end{equation}
Usually this $\delta(q)$ is defined for real $q$ and $k$, 
and it can diverge for complex $q$.

\begin{figure}[htb]
\begin{center}
\includegraphics[height=10cm]{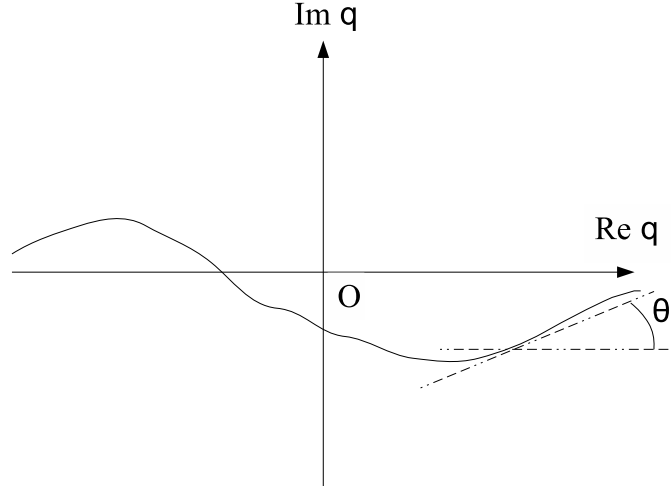}
\end{center}
\caption{An example of permitted paths}
\label{fig:contour}
\end{figure}

We now seek the possibility to define $\delta(q)$ for complex $q$ 
in the case that $k$ is also complex but 
the asymptotic value of $k$ is real. 
In this case we can take an arbitrary path running from 
$-\infty$ to $\infty$ in the complex plane of $k$. 
We call this path $C$ and define $\delta_c(q)$ and $\delta_c^\epsilon(q)$ for complex $q$ by  
\begin{eqnarray}
&&\delta_c(q) \equiv \lim_{\epsilon \rightarrow +0}  \delta_c^\epsilon(q) , \\
&&\delta_c^\epsilon(q) \equiv
\frac{1}{2\pi} \int_C e^{ikq - \epsilon k^2} dk 
= \sqrt{\frac{1}{4 \pi \epsilon}} e^{-\frac{q^2}{4\epsilon}} , \label{delta_c_epsilon(q)}
\end{eqnarray}
where we have introduced 
a finite but sufficiently small positive real number $\epsilon$, and in the second equality of eq.(\ref{delta_c_epsilon(q)}) 
we have assumed that $|k|$ goes larger than $\frac{1}{\sqrt{\epsilon}}$. 
We note that $e^{-\frac{q^2}{4\epsilon}}$ is convergent for $q$ such that 
\begin{equation}
L(q) \equiv \left( \text{Re}(q) \right)^2 - \left( \text{Im}(q) \right)^2 >0 . \label{cond_of_q_for_delta} 
\end{equation}
Since for any analytical test function $f(q)$ \footnote{Due to the Liouville theorem if $f$ is 
a bounded entire function, 
$f$ is constant. So we are considering as $f$ an unbounded entire function or a function 
which is not entire but is holomorphic at least in the region on which the path runs.} 
the path $C$ of the integral 
$\int_C f(q) e^{-\frac{q^2}{4\epsilon}} dq $ is independent of finite $\epsilon$, 
$\delta_c(q)$ satisfies for any $f(q)$ 
\begin{equation}
\int_{\text{along any permitted path from} -\infty ~ \text{to} ~+\infty } f(q) \delta_c(q) dq = f(0) ,
\end{equation}
as long as we choose a path such that at any $q$ 
its tangent line and a horizontal line 
form an angle $\theta$ whose absolute value  
is within $\frac{\pi}{4}$ to satisfy the inequality 
(\ref{cond_of_q_for_delta}).   
In fig.\ref{fig:contour} we have drawn one example of permitted paths. 

\begin{figure}[htb]
\begin{center}
\includegraphics[height=10cm]{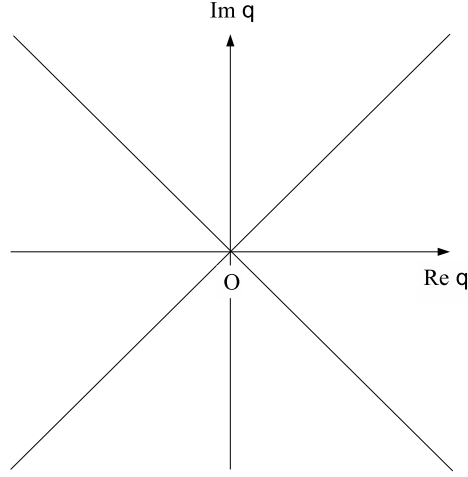}
\end{center}
\caption{Domain of the delta function}
\label{fig:delta_function}
\end{figure}

Thus we have extended the delta function to complex $q$ satisfying the condition 
(\ref{cond_of_q_for_delta}), and 
confirmed that it behaves as a distribution for any analytical test function $f(q)$. 
In fig.\ref{fig:delta_function} we have drawn the domain of the 
delta function. At the origin $\delta_c(q)$ is divergent. 
In the domain except for the origin, which is painted with inclined lines, 
$\delta_c(q)$ takes a vanishing value, 
while in the blank region the delta function is oscillating and divergent. 
$\delta_c(q)$ is well-defined for $q$ such that the condition 
(\ref{cond_of_q_for_delta}) is satisfied. 
%

\section{Explicit construction of $\hat{q}_{new}$, $\hat{p}_{new}$, $| q \rangle_{new}$ and $| p \rangle_{new}$ } 
\label{construction}
In this section 
we explicitly show one example of constructing the non-hermitian operators 
$\hat{q}_{new}$, $\hat{p}_{new}$, and the eigenstates of their hermitian conjugates 
$| q \rangle_{new}$ and $| p \rangle_{new}$ with complex eigenvalues $q$ and $p$, 
which satisfy eqs.(\ref{qhatqket=qqket_new})(\ref{phatpket=ppket_new})(\ref{commutator_q_p_new}), 
by squeezing coherent states so that we can have complex eigenvalues. 
Next we briefly explain the properties of $\hat{q}_{new}$, $\hat{p}_{new}$, $| q \rangle_{new}$ and $| p \rangle_{new}$ 
based on the analyses in appendix~\ref{explicitcalc}. 
Furthermore we also give a remark on an expression of a wave function $\psi(q)$.

\subsection{Definitions of $\hat{q}_{new}$, $\hat{p}_{new}$, $| q \rangle_{new}$ and $| p \rangle_{new}$}

We formally utilize coherent states $| \lambda \rangle_{coh}$ and $| \lambda' \rangle_{coh'}$ 
of two harmonic oscillators: one is defined with a mass $m$ and an angular frequency $\omega$, 
and the other is defined with $m'$ and $\omega'$. 
The explicit definition of $| \lambda \rangle_{coh}$ is given in appendix~\ref{app_cs}, 
and $| \lambda' \rangle_{coh'}$ is expressed similarly. 
Indeed, considering the following two relations, 
\begin{eqnarray}
&&\left( \hat{q} + i \frac{\hat{p}}{m\omega} \right) | \lambda \rangle_{coh} 
= \sqrt{\frac{2\hbar}{m\omega}} \lambda | \lambda \rangle_{coh} , \label{coh_q} \\
&&\left( \hat{p} + \frac{m' \omega'}{i} \hat{q} \right) | \lambda' \rangle_{coh'} = 
\frac{\lambda'}{i} \sqrt{2\hbar m' \omega'} | \lambda' \rangle_{coh'} , 
\label{coh_p}
\end{eqnarray}
we define $\hat{q}_{new}$ and $\hat{p}_{new}$ by 
\begin{eqnarray}
&&\hat{q}_{new} 
\equiv 
\frac{1}{ \sqrt{1 - \frac{m' \omega'}{m\omega} }  } \left( \hat{q} - i \frac{\hat{p}}{m\omega} \right), \label{def_qhat_new} \\
&&\hat{p}_{new} 
\equiv 
\frac{1}{ \sqrt{1 - \frac{m' \omega'}{m\omega} }  }  \left( \hat{p} - \frac{m' \omega'}{i} \hat{q} \right) , \label{def_phat_new} 
\end{eqnarray}
so that they obey eqs.(\ref{qhatqket=qqket_new})(\ref{phatpket=ppket_new})(\ref{commutator_q_p_new}). 
Also, we introduce $| q \rangle_{new}$ and $| p \rangle_{new}$ by 
\begin{eqnarray}
| q \rangle_{new} 
&\equiv& 
\left\{ \frac{m\omega}{4\pi \hbar} \left( 1 - \frac{m'\omega'}{m\omega} \right) \right\}^\frac{1}{4} 
e^{- \frac{m\omega}{4\hbar} \left( 1 - \frac{m'\omega'}{m\omega} \right) {q}^2 }
| \sqrt{ \frac{m\omega}{2\hbar} 
\left( 1 - \frac{m'\omega'}{m\omega} \right) } q \rangle_{coh} ,  \label{def_qketnew} \\
| p \rangle_{new} 
&\equiv& 
\left( \frac{1 - \frac{m'\omega'}{m\omega} }{4\pi \hbar m' \omega' } \right)^{\frac{1}{4}} 
e^{ -\frac{1}{4 \hbar m' \omega'}  \left( 1 - \frac{m'\omega'}{m\omega} \right)  p^2 }
| i \sqrt{ \frac{1}{2\hbar m' \omega'} 
\left( 1 - \frac{m'\omega'}{m\omega} \right)}  p \rangle_{coh'} . \label{def_pketnew}
\end{eqnarray}
The kets $| q \rangle_{new}$ and $| p \rangle_{new}$ are normalized by the following orthogonality relations, 
\begin{eqnarray}
{}_m\langle_{new} ~q' | q \rangle_{new} 
&=& \delta_c^{\epsilon_1} ( q' - q ) , \label{m_q'branew_qketnew}\\ 
{}_m\langle_{new} ~p' | p \rangle_{new} 
&=& \delta_c^{\epsilon'_1} ( p' - p ) ,  \label{m_p'branew_pketnew} 
\end{eqnarray}
where we have used the expression of eq.(\ref{delta_c_epsilon(q)}), and 
$\epsilon_1$ and $\epsilon'_1$ are given by 
\begin{eqnarray}
&&\epsilon_1 = \frac{\hbar}{m \omega \left( 1 - \frac{m'\omega'}{m\omega} \right)} , \label{epsilon_1} \\
&&\epsilon'_1 = \frac{\hbar m'\omega'}{1 - \frac{m'\omega'}{m\omega} } , \label{epsilon'_1}
\end{eqnarray}
with $m\omega$ and $m' \omega'$ taken sufficiently large and small respectively. 
Eq.(\ref{m_q'branew_qketnew}) is well defined for $q-q'$ satisfying 
the condition like eq.(\ref{cond_of_q_for_delta}), $L(q-q')>0$. 
We note that this condition is satisfied only when $q$ and $q'$ are on the same path. 
Similarly, eq.(\ref{m_p'branew_pketnew}) behaves well  
for complex $p$ and $p'$ such that $L(p-p')>0$. 
In the following we take $m\omega$ sufficiently large and $m'\omega'$ sufficiently small. 
As for the completeness relations for $|q \rangle_{new}$ and $|p \rangle_{new}$, 
appendix~\ref{completeness} confirms them as follows, 
\begin{eqnarray}
&&\int_C dq | q \rangle_{new} ~{}_m \langle_{new} ~q |  = 1 , \label{completion_complexq_ket2} \\
&&\int_C dp | p \rangle_{new} ~{}_m \langle_{new} ~p |  = 1 , \label{completion_complexp_ket2}
\end{eqnarray}
where $C$ is an arbitrary path running from $-\infty$ to $\infty$ in the complex plane of $q$ or $p$. 

To know more about $| q \rangle_{new}$ and $| p \rangle_{new}$ we first calculate 
$\langle q' | q \rangle_{new}$ and $\langle p' | q \rangle_{new}$ with real $q'$ and $p'$ 
by making use of eq.(\ref{q_rep_lambda_state}). 
They are expressed as 
\begin{eqnarray}
\langle q' | q \rangle_{new} 
&=& 
\sqrt{ \frac{m\omega}{2\pi \hbar} }
\left( 1 - \frac{m'\omega'}{m\omega} \right)^{\frac{1}{4}} 
e^{ \frac{p_0^2}{2 \hbar m\omega} } e^{ - \frac{i}{\hbar} q_0 p_0 } 
\exp\left[  -\frac{m\omega}{2\hbar} \left( q' - q_0 \right)^2 \right]  
e^{ \frac{i}{\hbar} p_0 q' } , \label{braq'qketnew2}    \nonumber \\
&& \\
\langle p' | q \rangle_{new} 
&=&  \frac{1}{ \sqrt{2\pi \hbar} } 
\left( 1 - \frac{m'\omega'}{m\omega} \right)^{\frac{1}{4}} 
e^{ \frac{p_0^2}{2 \hbar m\omega} } 
\exp\left[  -\frac{1}{2\hbar m\omega } \left( p' - p_0 \right)^2 \right] 
e^{ - \frac{i}{\hbar} q_0 p' } , 
\label{brap'qketnew2}
\end{eqnarray}
where $q_0$ and $p_0$ are written from eqs.(\ref{q_0})(\ref{p_0}) as 
\begin{eqnarray}
&&q_0 = \sqrt{ 1 - \frac{m'\omega'}{m\omega} } ~\text{Re} q  , \label{q_02} \\
&&p_0 = m \omega  \sqrt{ 1 - \frac{m'\omega'}{m\omega} }  ~\text{Im} q  . \label{p_02}
\end{eqnarray}
These expressions tell us that $|p_0|$ has to be taken large for large $m \omega$. 
Similarly, $\langle p' | p \rangle_{new}$ and $\langle q' | p \rangle_{new}$ with real $q'$ and $p'$ 
are estimated by utilizing eq.(\ref{p_rep_lambda_state}) as  
\begin{eqnarray}
\langle p' | p \rangle_{new} 
&=& 
\frac{1}{ \sqrt{ 2\pi \hbar m' \omega' } }
\left( 1 - \frac{m'\omega'}{m\omega}  \right)^{\frac{1}{4}} 
e^{ \frac{ m' \omega' }{2 \hbar} {q'_0}^2 } e^{ \frac{i}{\hbar} q'_0 p'_0 } 
\exp\left[  -\frac{1}{2\hbar m' \omega'} \left( p' - p'_0 \right)^2 \right]  
e^{ -\frac{i}{\hbar} q'_0 p' } , \nonumber \\
&& \label{brap'pketnew2} \\
\langle q' | p \rangle_{new} 
&=& 
\frac{1}{ \sqrt{ 2\pi \hbar } }
\left( 1 - \frac{m'\omega'}{m\omega}  \right)^{\frac{1}{4}} 
e^{ \frac{ m' \omega' }{2 \hbar} {q'_0}^2 } 
\exp\left[  -\frac{m' \omega'}{2\hbar } \left( q' - q'_0 \right)^2 \right]  
e^{ \frac{i}{\hbar} p'_0 q' } , \label{braq'pketnew2}
\end{eqnarray}
where $q'_0$ and $p'_0$ are expressed from eqs.(\ref{q_0})(\ref{p_0}) as 
\begin{eqnarray}
&&q'_0 =\frac{-1}{m' \omega'}  \sqrt{ 1 - \frac{m'\omega'}{m\omega} }  ~\text{Im} p  , \label{q_03} \\
&&p'_0 =  \sqrt{ 1 - \frac{m'\omega'}{m\omega} } ~\text{Re} p . \label{p_03}
\end{eqnarray}
These expressions show us that $|q'_0|$ has to be taken large for small $m' \omega'$.

\subsection{Properties of $\hat{q}_{new}^\dag$, $\hat{p}_{new}^\dag$, $| q \rangle_{new}$ and $| p \rangle_{new}$}\label{propqketnewpketnew}

The operators $\hat{q}_{new}^\dag$ and $\hat{p}_{new}^\dag$ behave like 
$\hat{q}$ and $\hat{p}$ respectively for  $| q' \rangle$ with real $q'$ or $| p' \rangle$ with real $p'$, 
as studied in appendix~\ref{qpnewhatreal}. 
Also, the kets $| q' \rangle_{new}$ with real $q'$ and $| p' \rangle_{new}$ with real $p'$ become 
$| q' \rangle$ and $| p' \rangle$ respectively, 
as seen in appendix~\ref{q'ketp'ket}. 
Appendix~\ref{phatqket_qhatpket} confirms the following relations, 
\begin{eqnarray}
\hat{p}_{new}^\dag| q \rangle_{new} 
&\simeq&i \hbar \frac{\partial}{\partial q} | q \rangle_{new} \quad \text{for small $m' \omega'$} , \label{phatnewqketnew2} \\
\hat{q}_{new}^\dag| p \rangle_{new} 
&\simeq& \frac{\hbar}{i} \frac{\partial}{\partial p} | p \rangle_{new} \quad \text{for large $m \omega$} , 
\label{qhatnewpketnew2}
\end{eqnarray}
which are similar to eqs.(\ref{phatqket=ihbardeldelqqket})(\ref{qhatpket=ihbardeldelppket}) respectively. 
Furthermore we have 
\begin{equation} 
{}_m\langle_{new} ~q | p \rangle_{new} 
\simeq \frac{1}{\sqrt{2 \pi \hbar}} e^{\frac{i}{\hbar}p q}  
\quad \text{for large $m \omega$ and small $m' \omega'$} , \label{basis_fourier_transf2}  
\end{equation}
which stands for any $q$ and $p$ regardless of complex or real, as studied in appendix~\ref{basisfourier}. 
These relations show that $| q \rangle_{new}$, $| p \rangle_{new}$, $\hat{q}_{new}^\dag$ and $\hat{p}_{new}^\dag$ 
obey the same relations as $| q \rangle$, $| p \rangle$, $\hat{q}$ and $\hat{p}$ satisfy.

\subsection{A remark on an expression of a wave function $\psi(q)$}\label{remark_wavefunc}

Before ending this section we give a remark on an expression of a wave function $\psi(q)$.  
In the RAT it is expressed in terms of bras and kets as 
\begin{equation}
\psi(q) = \langle q | \psi \rangle , 
\end{equation}
and it cannot be used for complex $q$ because $\langle q |$ is defined only for real $q$. 
On the other hand, in our formalism even for complex $q$ we can express it as
\begin{equation} 
\psi(q) = ~{}_m\langle_{new}~ q | \psi \rangle . 
\end{equation}
This is an explicit representation of the analytically extended wave function in terms of bras and kets. 
Indeed this becomes the usual one $\langle q | \psi \rangle$ for real $q$. 
Only in our formalism can we express it for complex $q$ besides real $q$. 
Also, we note that in our formalism $\psi(q)$ makes up the Hilbert space with the following norm 
\begin{equation}
\int_C \psi(q)^{*_q}  \psi(q) dq < \infty , 
\end{equation} 
where $C$ is any path running from $-\infty$ to $\infty$ in the complex plane of $q$. 
Thus $\psi(q)$ is normalized by 
\begin{equation}
\int_C \psi(q)^{*_q} \psi(q) dq = 1 . 
\end{equation}
So $\psi(q)^{*_q} \psi(q)$, which becomes $| \psi(q) |^2$ for real $q$, looks like a probability density defined 
on $C$. 
But for complex $q$ it is not real, so we cannot interpret it as a true probability density; it is a formal one.

Finally, for our convenience we show the summary of the comparison of the RAT and the CAT 
in table \ref{tab1:comparison}. 
Furthermore, in appendix~\ref{theorem}, 
to make clear the relation between functions on the phase space describing some classical variables 
and the corresponding operators under quantization, 
providing the notions of ``$\epsilon$-analytical" functions, 
and ``expandable" and  ``$\epsilon$-expandable" operators, we pose a theorem, 
which claims that if and only if some operator corresponding to an $\epsilon$-expandable 
function on the phase space, its matrix element in $q$-representation is an $\epsilon$-analytical function. 

\begin{table}
\caption[AAA]{Various quantities in the RAT and the CAT}
\label{tab1:comparison}
\begin{center}
\begin{tabular}{|p{4cm}|p{3.5cm}|p{5.5cm}|}
\hline
    & the RAT & the CAT \\
\hline
parameters  &    $q$, $p$ real,   & $q$, $p$ complex \\  
\hline
complex conjugate  &    $*$   &  $*_{ \{~\}}$  \\  
\hline
hermitian conjugate  &    $\dag$   &  $\dag_{ \{~\}}$  \\  
\hline
delta function of $q$ &  $\delta(q)$ defined for    & $\delta_c(q)$ defined for $q$ s.t.\\
& real $q$ &  $\left( \text{Re}(q) \right)^2 > \left( \text{Im}(q) \right)^2$  \\ 
\hline
bras of $| q\rangle$, $| p\rangle$   &  $\langle q | = ( | q\rangle )^\dag $,  & 
${}_m\langle_{new}~ q |=\langle_{new}~ q^* |  = ( | q \rangle_{new} )^{\dag_q}  $,  \\
& $\langle p | = ( | p\rangle )^\dag$ & ${}_m\langle_{new}~ p |=\langle_{new}~ p^* |  = ( | p \rangle_{new} )^{\dag_p}  $ \\

\hline  
completeness for  &  $\int_{-\infty}^\infty |q\rangle \langle q | dq =1$ , & 
$\int_C |q \rangle_{new} ~{}_m\langle_{new}~ q | dq =1$ , \\
$| q \rangle$ and $|p\rangle$ &  $\int_{-\infty}^\infty |p \rangle \langle p | dp =1$  & 
$\int_C |p\rangle_{new} ~{}_m\langle_{new}~ p | dp =1$ \\
& along real axis & $C$: any path running from $-\infty$ to $\infty$\\
\hline 
orthogonality for  &  $\langle q | q'\rangle = \delta(q-q')$ , & 
${}_m\langle_{new}~ q | q'\rangle_{new} = \delta_c^{\epsilon_1} (q-q')$ , \\
$| q\rangle$ and $|p\rangle$ &  $\langle p | p'\rangle = \delta(p-p')$ & 
${}_m\langle_{new}~ p | p'\rangle_{new} = \delta_c^{\epsilon'_1} (p-p')$ \\
\hline 
basis of Fourier expansion &  $\langle q | p\rangle = \exp(ipq)$ & 
${}_m\langle_{new}~ q | p\rangle_{new} = \exp(ipq)$ \\
\hline 
$q$ representation of  $| \psi \rangle$  & $\psi(q) = \langle q | \psi \rangle$ & 
$\psi(q) = {}_m\langle_{new}~ q | \psi \rangle $ \\
\hline 
complex conjugate of  $\psi(q)$ & $\langle q | \psi \rangle^*=\langle \psi | q \rangle$ & 
${}_m\langle_{new}~ q | \psi \rangle^{*_q } =\langle \psi | q \rangle_{new}$ \\
\hline
normalization of $\psi(q)$  & $\int_{-\infty}^\infty \psi(q)^* \psi(q) dq = 1$ & 
$\int_C \psi(q)^{*_q } \psi(q) dq = 1$ \\
\hline
\end{tabular}
\end{center}
\end{table}
%

\section{A possible mechanism to obtain a hermitian Hamiltonian}\label{possible_mechanism}

Hamiltonians, which are correlated to complex actions, are non-hermitian. 
Recently, a class of non-hermitian Hamiltonians 
satisfying the PT symmetry has been intensively studied 
in various directions.\cite{PTsym_Hamiltonians}\cite{Experiments}\cite{Erdem:2006zh} 
The eigenvalues are real, and thus the Hamiltonians 
give consistent quantum theories. 
In ref.\cite{Nagao:2010xu}, on the other hand, 
we have studied the time-development of some state in a system defined by 
a generic non-hermitian diagonalizable bounded Hamiltonian, 
and have presented a possible mechanism to obtain a hermitian Hamiltonian 
in the real coordinate case. 
In this section, as an application of the complex formalism which we have developed in the foregoing sections, 
we attempt to extend the mechanism proposed in ref.\cite{Nagao:2010xu} 
to the complex coordinate formalism 
by utilizing the philosophy of keeping the analyticity in path integral variables and the devices 
such as modified bras. 
We begin with the discussion of the difficulty due to the non-hermiticity of the Hamiltonian.

\subsection{Difficulties with the non-hermitian Hamiltonian}

If we naively define a time-development operator from the time $t_0$ to $t$ by 
\begin{equation}
U_{t_0 \rightarrow t} =\exp\left(-\frac{i}{\hbar}H(t -t_0) \right) ,
\end{equation}
$U_{t_0 \rightarrow t}$ is not unitary. 
This is a big problem, and it sounds excluded as a model that could be expected 
to be realized in nature. 
Indeed if we start by a state $| \psi(t_0) \rangle$ at time $t_0$
and develop it by $U_{t_0 \rightarrow t} $ into a result state $| \psi(t) \rangle$ at 
time $t$, which is given by 
\begin{equation}
| \psi(t) \rangle= U_{t_0 \rightarrow t} | \psi(t_0) \rangle ,
\end{equation}
then we encounter the non-conservation of its probability 
\begin{equation}
| \langle \psi(t) | \psi(t) \rangle |^2 \neq | \langle \psi(t_0) | \psi(t_0) \rangle |^2.
\end{equation}

Supposedly one should make an interpretation that would correspond to
normalizing the wave function coming out of a time-development by means of
the non-hermitian Hamiltonian $H$. 
In order to get a reasonable interpretation 
we could decide to rescale the resulting state $| \psi(t) \rangle$ by simply normalizing it. 
As done in ref.\cite{Nielsen:2007mj}, we replace it by 
\begin{equation}
|  \psi(t) \rangle_N = \frac{1}{\sqrt{ \langle \psi(t) |  \psi(t) \rangle}} |  \psi(t) \rangle . \label{psiNket}
\end{equation}
Then we do not have at least any probability for the world stopping to exist or getting multiplied. 
However, since the normalization factor depends on the time $t$, 
$|  \psi(t) \rangle_N$ satisfies the slightly modified Schr\"{o}dinger equation, 
\begin{eqnarray}
i\hbar \frac{\partial}{\partial t} |  \psi(t) \rangle_N 
&=& H | \psi(t) \rangle_N - ~{}_N\langle \psi(t) | \frac{ H-H^\dag }{2}  
| \psi(t) \rangle_N | \psi(t) \rangle_N \nonumber \\ 
&=& H_h | \psi(t) \rangle_N 
+ \left\{ H_a  - ~{}_N\langle \psi(t) | H_a | \psi(t) \rangle_N \right\} | \psi(t) \rangle_N . 
\label{psiN_mod_schr}
\end{eqnarray}
Also, if we define the expectation value of some operator ${\cal O}$ by 
\begin{equation}
\bar{\cal O}(t) \equiv  {}_{N} \langle \psi(t) | {\cal O} | \psi(t) \rangle_{N} 
=  {}_{N} \langle \psi(t_0) | {\cal O}_{H}(t, t_0) | \psi(t_0) \rangle_{N} , 
\end{equation} 
where we have introduced the time-dependent operator in the Heisenberg picture, 
\begin{equation}
{\cal O}_{H}(t, t_0) \equiv \frac{ \langle \psi(t_0) | \psi(t_0) \rangle }{ \langle \psi(t) | \psi(t) \rangle } 
e^{ \frac{i}{\hbar} H^{\dag} (t-t_0) } {\cal O} e^{ -\frac{i}{\hbar} H(t-t_0) } , 
\end{equation}
we see that ${\cal O}_{H}(t, t_0)$ obeys the slightly modified Heisenberg equation, 
\begin{eqnarray}
i\hbar \frac{d}{dt} {\cal O}_{H}(t, t_0)  
&=& {\cal O}_{H}(t, t_0) H - H^{\dag} {\cal O}_{H}(t, t_0) 
-2 {}_{N} \langle \psi(t) | H_{a} | \psi(t) \rangle_{N}  {\cal O}_{H}(t, t_0) \nonumber \\
&=& [ {\cal O}_{H}(t, t_0) , H_{h} ] 
+ \left\{  {\cal O}_{H}(t, t_0) , H_{a} -  {}_{N} \langle \psi(t) | H_{a} | \psi(t) \rangle_{N} \right\} . \label{psiN_mod_heisen}
\end{eqnarray}
Eq.(\ref{psiN_mod_schr}) shows that the anti-hermitian part of the Hamiltonian $H_a$ 
is considerably suppressed in the classical limit, but 
the effect cannot be removed completely at the quantum level 
even if we choose the basis so that $H_a$ is diagonalized, 
since eq.(\ref{psiN_mod_schr}) is non-linear with regard to $| \psi(t) \rangle_N$. 
Such an effect of $H_a$ exists also in (\ref{psiN_mod_heisen}).

Besides the above difficulty we know that the eigenvalues of the non-hermitian Hamiltonian 
are not real in general. 
Furthermore, since the eigenstates are not orthogonal, 
a transition that should not be possible could be measured. 
From these properties it does not look a physically reasonable theory, 
but in ref.\cite{Nagao:2010xu} we have shown a possible way to circumvent this problem 
in the real coordinate case via two steps based on the speculation in ref.\cite{originsym}. 
We shall show that the two steps can be applied also in the complex coordinate formalism. 
The first step is to define a physically reasonable inner product $I_Q$ such that 
the eigenstates of the Hamiltonian get orthogonal with regard to it, and thus 
it gives us a true probability for a transition from some state to another. 
As we shall see later, $I_Q$ makes the Hamiltonian normal with regard to it. 
In other words $I_Q$ has to be defined for consistency so that the Hamiltonian 
- even if it cannot be made hermitian - at least be normal.  
We explain how a reasonable physical assumption about the probabilities leads to 
the proper inner product $I_Q$, 
and define a hermiticity with regard to $I_Q$, $Q$-hermiticity. 
The second step is to use a mechanism of 
suppressing the effect of the anti-hermitian part of the Hamiltonian $H$ 
after a long time development. 
We shall explicitly show the mechanism 
with the help of the proper inner product $I_Q$.  
For the states with high imaginary part of eigenvalues of $H$, 
the factor $\exp\left(-\frac{i}{\hbar} H(t - t_0) \right)$ will grow exponentially with $t$. 
After a long time the states with the highest imaginary part of eigenvalues of $H$ 
get more favored to result than others. 
Thus the effect of the imaginary part gets attenuated. 
Utilizing this effect to normalize the state, 
we can effectively obtain a $Q$-hermitian Hamiltonian. 
%

\subsection{Physical significance of an inner product} 

The Born rule of quantum mechanics is well-known in the form: 
When a quantum mechanical system prepared in a state $| i \rangle$ 
at time $t_i$ time-develops into 
$| i (t_f) \rangle = e^{-\frac{i}{\hbar}H(t_f -t_i)}| i \rangle$ 
at time $t_f$, 
we will measure it in a state $|f \rangle$ with the probability 
$P_{f ~\text{from} ~i} = | \langle f | i(t_f) \rangle |^2$. 
We note that the probability depends on how we define an inner product 
of the Hilbert space. 
A usual inner product is defined as a sesquilinear form. 
We denote it as 
$I( | f \rangle, | i(t_f) \rangle )= \langle f | i(t_f) \rangle$. 
It is $|I( | f \rangle, | i(t_f) \rangle )|^2$ that we measure 
by seeing how often we get $| f \rangle $ from $| i(t_f) \rangle $. 
Measuring the transition of superposition like $c_1 | a \rangle + c_2 | b \rangle$ repeatedly, 
we can extract the whole form of $I(| f \rangle, | i(t_f) \rangle)$ 
of any two states by using the sesquilinearity.

To consider an inner product in our theory with non-hermitian Hamiltonian $H$, 
we assume that $H$ is diagonalizable, and diagonalize $H$ by using a non-unitary operator $P$ as 
\begin{equation}
H = PD P^{-1}. 
\end{equation}
We introduce an orthonormal basis $| e_i \rangle (i=1, \ldots)$ satisfying 
$\langle e_i | e_j \rangle = \delta_{ij}$ by 
$D | e_i \rangle = \lambda_i   | e_i \rangle$, 
where $\lambda_i (i=1, \ldots)$ are generally complex.
We also introduce the eigenstates $| \lambda_i \rangle$ of $H$ by 
$| \lambda_i \rangle = P | e_i \rangle$, which obeys 
\begin{equation}
H | \lambda_i \rangle = \lambda_i | \lambda_i \rangle.
\end{equation} 
We note that $| \lambda_i \rangle$ are not orthogonal to each other in the usual inner product $I$, 
$\langle \lambda_i | \lambda_j \rangle \neq \delta_{ij}$.

As we are prepared, let us apply the usual inner product $I$ to our theory 
with the non-hermitian Hamiltonian $H$, 
and consider a transition from an eigenstate $| \lambda_i \rangle$ 
to another $| \lambda_j \rangle ~(i \neq j)$ fast in time $\Delta t$. 
Then, though $H$ cannot bring the system from one eigenstate $| \lambda_i \rangle$ 
to another one $| \lambda_j \rangle ~(i \neq j)$,  
the transition can be measured, that is to say, 
$|I(| \lambda_j \rangle, \exp\left( -\frac{i}{\hbar} H \Delta t \right) |\lambda_i \rangle )|^2 \neq 0$, 
since the two eigenstates are not orthogonal to each other. 
Such a transition should be prohibited in a reasonable theory, 
based on the philosophy that a measurement - even performed in a short time - is fundamentally 
a physical development in time. 
Thus we think that the eigenstates have to be orthogonal to each other.

\subsection{A proper inner product and hermitian conjugate}

Since we are physically entitled to require that a truly functioning measurement procedure 
must necessarily have reasonable probabilistic results, 
for arbitrary states $| \psi_1 \rangle$ and $|\psi_2 \rangle$ 
we attempt to construct 
a proper inner product 
\begin{equation}
I_Q(|\psi_2 \rangle,|\psi_1 \rangle)=\langle \psi_2 |_Q \psi_1 \rangle \equiv \langle \psi_2 | Q | \psi_1 \rangle ,
\end{equation}
with the property that the eigenstates $| \lambda_i \rangle$ and $| \lambda_j \rangle$ get orthogonal to each other, 
\begin{equation}
I_Q( | \lambda_i \rangle, | \lambda_j \rangle ) = \delta_{ij} , \label{IQ_delta_ij}
\end{equation}
where $Q$ is some operator chosen appropriately. 
Of course in the special case of the Hamiltonian $H$ being hermitian $Q$ would be 
the unit operator. 
We believe that the true probability is given by such a proper inner product $I_Q$, 
based on which 
the Hamiltonian is conserved even if it is not hermitian and typically has complex eigenvalues. 
This condition applies to not only the eigenstates of the Hamiltonian 
but also those of any other conserved quantities. 
The transition from an eigenstate of such a conserved quantity to another eigenstate 
with a different eigenvalue should be prohibited in a reasonable theory. 
Furthermore for the case where $|\psi_2 \rangle$ is given in a parametrized state $|u \rangle$,  
we define another proper inner product with a modified bra by 
\begin{equation}
I_Q^m(| u \rangle, | \psi_1 \rangle) \equiv {}_m\langle u |_Q \psi_1 \rangle 
= {}_m\langle u | Q | \psi_1 \rangle .
\end{equation}
This is expected to be used for the purpose of keeping the analyticity in $u$.

In the RAT the usual inner product $I$ is defined to satisfy 
$\langle \psi_1(t) | \psi_2(t) \rangle = \langle \psi_2(t) | \psi_1(t) \rangle^*$. 
Hence we impose a similar relation on $I_Q$ as 
\begin{equation}
\langle \psi_1(t) |_Q \psi_2(t) \rangle = \langle \psi_2(t) |_Q \psi_1(t) \rangle^* . 
\label{comp_conj_prop_inner_prod}
\end{equation} 
Then we obtain a condition 
\begin{equation}
Q^\dag=Q, \label{def_Q_hermiticity1}
\end{equation}
namely, $Q$ has to be hermitian. 
In the case where $|\psi_1 \rangle$ or $|\psi_2 \rangle$ are given in 
parametrized states as $|u \rangle$ or $|v \rangle$, 
we extend the condition of eq.(\ref{comp_conj_prop_inner_prod}) to 
\begin{equation}
{}_{\{ \}}\langle u |_Q v \rangle = {}_{\{ \}}\langle v |_Q u \rangle^{*_{\{ \}}} ,   
\label{comp_conj_prop_inner_prod5}
\end{equation} 
and we have the same relation as eq.(\ref{def_Q_hermiticity1}). 
We choose the set of parameters $\{ \}$ 
according to which parameters we want to keep the analyticity in.

Via the inner product $I_Q$, we define the corresponding hermitian conjugate $\dag^Q$ 
for some operator $A$ by 
\begin{equation}
\langle \psi_2 |_Q A | \psi_1 \rangle^* = \langle \psi_1 |_Q A^{\dag^Q} | \psi_2 \rangle , \label{def_dag_n}
\end{equation}
from which we obtain 
\begin{equation}
A^{\dag^Q} = Q^{-1} A^\dag Q . \label{def_A_dag_Q_1}
\end{equation} 
Similarly, in the case where $|\psi_1 \rangle$ or $|\psi_2 \rangle$ are given in states 
as $|u \rangle$ or $|v \rangle$, we extend eq.(\ref{def_dag_n}) to 
\begin{equation}
{}_{\{ \}}\langle v |_Q A | u \rangle^{*_{\{ \}}} = 
{}_{\{ \}}\langle u |_Q A^{\dag^Q} | v \rangle , 
\label{def_dag_n5}
\end{equation} 
and have the following relation,
\begin{equation}
A^{\dag^Q_{\{\}} } = A^{\dag^Q} . \label{herm_conj_op2Q}
\end{equation}
We also define ${\dag^Q}$ for kets and bras as 
$| \lambda \rangle^{\dag^Q} \equiv \langle \lambda |_Q $ and 
$\left(\langle \lambda |_Q \right)^{\dag^Q} \equiv | \lambda \rangle$ 
so that we can manipulate $\dag^Q$ like a usual hermitian conjugate $\dag$. 
Similarly, we define $\dag^Q_{\{ \}}$ for kets and bras as 
$| \lambda \rangle^{\dag^Q_{ \{ \} }} \equiv {}_{ \{ \} }\langle \lambda |_Q $ and 
$\left( {}_{ \{ \} } \langle \lambda |_Q \right)^{\dag^Q_{ \{ \} }} \equiv | \lambda \rangle$.

Furthermore we define a hermiticity with regard to the new inner product. 
When $A$ satisfies 
\begin{equation}
A^{\dag^Q} = A ,  
\end{equation}
we call $A$ $Q$-hermitian. 
This is the definition of the $Q$-hermiticity. 
Since this relation can be expressed as $Q A = ( Q A )^\dag$, 
when $A$ is $Q$-hermitian, $Q A$ is hermitian, 
and vice versa.\footnote{We note that in ref.\cite{Geyer} a similar inner product 
has been studied and a criterion for identifying a unique
inner product through the choice of physical observables has been also provided.}

If some operator $A$ can be diagonalized as $A=P_A D_A P_A^{-1}$, 
then $Q$-hermitian conjugate of $A$ is expressed as  
\begin{equation}
A^{\dag^Q} = Q^{-1} (P_A D_A P_A^{-1})^{\dag} Q , 
\end{equation}
where we have used eq.(\ref{def_A_dag_Q_1}). 
If we choose $Q$ as $Q= (P_A^\dag)^{-1} P_A^{-1}$, 
which satisfies $Q^\dag=Q$, 
we have $A^{\dag^Q}  = P_A D_A^\dag P_A^{-1}$. 
Therefore, if $D_A$ satisfies $D_A^\dag=D_A$, 
which means that the diagonal components of $D_A$ are real,  
then $A$ is shown to be $Q$-hermitian. 
In the following we define $Q$ by 
\begin{equation}
Q= (P^\dag)^{-1} P^{-1} \label{def_Q_PdaginPin}
\end{equation} 
with the diagonalizing matrix $P$ of the non-hermitian Hamiltonian $H$. 
We note that $I_Q$ or $I_Q^m$ are different from the CPT inner product defined 
in the PT symmetric Hamiltonian formalism\cite{PTsym_Hamiltonians}.

\subsection{Normality of the Hamiltonian}

To prove that the non-hermitian Hamiltonian $H$ is $Q$-normal, 
i.e. normal with regard to the inner product $I_Q$, we first define 
\begin{equation}
``P^{\dag^Q}"
\equiv
\left(
 \begin{array}{c}
      \langle \lambda_1 |_Q     \\
      \langle \lambda_2 |_Q     \\
      \vdots 
 \end{array}
\right) 
\end{equation}
by using the diagonalizing operator $P$ of $H$, which has a structure as 
$P=( |\lambda_1 \rangle, |\lambda_2 \rangle, \ldots  )$, where $| \lambda_i \rangle$ are 
eigenstates of $H$. 
We note that $``P^{\dag^Q}"$ is defined 
by using the $Q$-hermitian conjugate of kets, 
so $``P^{\dag^Q}" \neq Q^{-1} P^\dag Q$. 
Then since $\langle \lambda_i |_Q \lambda_j \rangle = \delta_{ij}$, 
we see that $``P^{\dag^Q}" P = {\mathbf 1}$, namely, $``P^{\dag^Q}"=P^{-1} $. 
Hence we can say that $P$ is $Q$-unitary.

Next we consider the relation $``P^{\dag^Q}" H P = D $. 
The $(i,j)$-component of this relation in $|\lambda_i  \rangle$ basis is written as 
$\langle \lambda_i |_Q H | \lambda_j \rangle = \lambda_i \delta_{ij} $. 
Taking the complex conjugate, 
we obtain $\langle \lambda_j |_Q H^{\dag^Q} | \lambda_i \rangle = 
\lambda_i^* \delta_{ij} $, 
that is to say, $\langle \lambda_i |_Q H^{\dag^Q} | \lambda_j \rangle = \lambda_i^* \delta_{ij} $. 
This is written in the operator form as 
$``P^{\dag^Q}" H^{\dag^Q} P = D^{\dag} $. 
Therefore we obtain 
\begin{equation}
[H, H^{\dag^Q} ] = P [D, D^\dag ] P^{-1} =0.
\end{equation} 
Thus we see that $H$ is $Q$-normal. 
In other words we can say that the inner product $I_Q$ is defined so that 
$H$ is normal with regard to it.

Furthermore for later convenience we decompose $H$ as 
\begin{equation}
H=H_{Qh} + H_{Qa},
\end{equation}
where $H_{Qh}$ and $H_{Qa}$ are defined by 
\begin{eqnarray}
&&H_{Qh} = \frac{H + H^{\dag^Q} }{2} , \\ 
&&H_{Qa} = \frac{H - H^{\dag^Q} }{2} .
\end{eqnarray}
They are $Q$-hermitian and anti-$Q$-hermitian parts of $H$ respectively. 
We also decompose $D$ as 
\begin{equation}
D=D_R + iD_I, 
\end{equation}
where $D_R$ and $D_I$ are defined by 
\begin{eqnarray}
&&D_R= \frac{D + D^\dag }{2} , \\
&&D_I= \frac{D - D^\dag }{2} . 
\end{eqnarray}
The diagonal components of $D_R$ and $D_I$ are the real 
and imaginary parts of 
the diagonal components of $D$ respectively. 
Then $H_{Qh}$ and $H_{Qa}$ can be expressed in terms of $D_R$ and $D_I$ as 
\begin{eqnarray}
&&H_{Qh} = P D_R P^{-1}, \\
&&H_{Qa} = i P D_I P^{-1}. \label{HQa_paren}
\end{eqnarray}

\subsection{Normalization of $| \psi \rangle$ and expectation value}

We consider some state $| \psi(t) \rangle $, which obeys 
the Schr\"{o}dinger equation 
$i \hbar \frac{d}{dt} | \psi (t) \rangle = H | \psi (t) \rangle $. 
Normalizing it as 
\begin{equation}
 |\psi(t) \rangle_{N} \equiv \frac{1}{\sqrt{ \langle {\psi}(t) |_Q ~{\psi}(t) \rangle} } | {\psi}(t) \rangle ,
\end{equation} 
we define the expectation value of some operator ${\cal O}$ by 
\begin{equation}
\bar{\cal O}_Q(t) \equiv  {}_{N} \langle \psi(t) |_Q {\cal O} | \psi(t) \rangle_{N} 
=  {}_{N} \langle \psi(t_0) |_Q {\cal O}_{QH}(t, t_0) | \psi(t_0) \rangle_{N},
\end{equation} 
where ${}_{N} \langle \psi(t) |$ is given by 
\begin{equation}
{}_{N} \langle \psi(t) | \equiv ( |\psi(t) \rangle_{N} )^{\dag}
=\frac{1}{\sqrt{ \langle {\psi}(t) |_Q ~{\psi}(t) \rangle} } \langle \psi(t) | ,
\end{equation}
and we have introduced the time-dependent operator in the Heisenberg picture, 
\begin{equation}
{\cal O}_{QH}(t ,t_0) \equiv \frac{ \langle \psi(t_0) |_Q \psi(t_0) \rangle }{ \langle \psi(t) |_Q \psi(t) \rangle } 
e^{ \frac{i}{\hbar} H^{\dag^Q} (t-t_0) } {\cal O} e^{ -\frac{i}{\hbar} H(t-t_0) }.
\end{equation}

Since the normalization factor depends on time $t$, $| \psi(t) \rangle_{N} $ does not 
obey the Schr\"{o}dinger equation, but rather the slightly modified Schr\"{o}dinger equation,
\begin{eqnarray}
i\hbar \frac{d}{dt} | \psi(t) \rangle_{N} 
&=& H | \psi(t) \rangle_{N} 
-{}_{ N} \langle \psi(t) |_Q H_{Qa} | \psi(t) \rangle_{N} | \psi(t) \rangle_{N} \nonumber \\
&=& H_{Qh} | \psi(t) \rangle_{N} 
+ \left( H_{Qa} -{}_{N} \langle \psi(t) |_Q H_{Qa} | \psi(t) \rangle_{N} \right) | \psi(t) \rangle_{N} . \label{sch}
\end{eqnarray} 
In addition the time development equation for ${\cal O}_{QH}(t , t_0)$ 
is seen to be 
\begin{eqnarray}
&& i\hbar \frac{d}{dt} {\cal O}_{QH}(t , t_0)  \nonumber \\
&=& {\cal O}_{QH}(t, t_0) H - H^{\dag^Q} {\cal O}_{QH}(t, t_0) 
-2 {}_{N} \langle \psi(t) |_Q H_{Qa} | \psi(t) \rangle_{N}  {\cal O}_{QH}(t, t_0) \nonumber \\
&=& [ {\cal O}_{QH}(t, t_0) , H_{Qh} ] 
+ \left\{  {\cal O}_{QH}(t, t_0) , H_{Qa} -  {}_{N} \langle \psi(t) |_Q H_{Qa} | \psi(t) \rangle_{N} \right\} . \label{hei}
\end{eqnarray}
This is the slightly modified Heisenberg equation.

In eqs.(\ref{sch})(\ref{hei}) we find the effect of $H_{Qa}$, the anti-$Q$-hermitian part of the Hamiltonian $H$, 
though it seems to disappear in the classical limit. 
It is intriguing that in that limit eqs.(\ref{sch})(\ref{hei}) are expressed as 
\begin{eqnarray}
&&i\hbar \frac{d}{dt} | \psi(t) \rangle_{N} \simeq H | \psi(t) \rangle_{N} , \\
&&i\hbar \frac{d}{dt} {\cal O}_{QH}(t, t_0) \simeq [ {\cal O}_{QH}(t, t_0) , H_{Qh} ] .
\end{eqnarray} 
On the other hand, with the second step we explain next, 
we shall find that in both of the equations the effect of $H_{Qa}$ disappears at the quantum level.

\subsection{The mechanism for suppressing the anti-$Q$-hermitian part of the Hamiltonian}

To show the mechanism for suppressing the effect of $H_{Qa}$, 
we study the time-development of $| \psi(t) \rangle$ explicitly. 
We introduce $| \psi'(t) \rangle $ by 
$| \psi' (t) \rangle = P^{-1} | \psi (t) \rangle$, and expand it as 
$| \psi'(t) \rangle = \sum_i a_i(t) | e_i \rangle $. 
Then $| \psi(t) \rangle $ can be written in an expanded form as 
$| \psi(t) \rangle = \sum_i a_i(t) | \lambda_i \rangle $. 
Since $ | \psi'(t) \rangle$ obeys 
$i \hbar \frac{d}{dt} | \psi'(t) \rangle =D | \psi'(t) \rangle$, 
the time-development of $| \psi(t) \rangle$ from some time $t_0$ is calculated as 
\begin{eqnarray}
| \psi(t) \rangle &=& P e^{- \frac{i}{\hbar} D (t-t_0)} | \psi'(t_0) \rangle \nonumber\\
&=& \sum_i a_i(t_0) 
e^{ \frac{1}{\hbar} \left( \text{Im} \lambda_i - i \text{Re} \lambda_i \right) (t-t_0)}       
| \lambda_i \rangle . \
\end{eqnarray}

$\text{Im} \lambda_i $ is related to the anti-$Q$-hermitian part of the Hamiltonian, 
$H_{Qa}$, as seen from eq.(\ref{HQa_paren}). 
Now we assume the boundedness of $H$. 
Then we can crudely imagine that some of $\text{Im} \lambda_i $ 
take the maximal value $B$. 
We denote the corresponding subset of $\{ i \}$ as $A$. 
If we imagine a classical approximation, we can consider theTaylor-expansion of $H_{Qa}$ around the value $B$. 
Thus we get a good approximation to the practical outcome of the model. 
In the Taylor-expansion we do not have the linear term because we expand it near the maximum, 
so we get only non-trivial terms of {\em second order}. 
In this way $H_{Qa}$ becomes constant in the first approximation, 
and thus it is not so important observationally.
Therefore, if a long time has passed, namely for large $t-t_0$, 
the states with $\text{Im} \lambda_i |_{i \in A}$ survive and contribute most in the sum.

To see how $| \psi(t) \rangle $ is effectively described for large $t-t_0$, 
we introduce a diagonalized Hamiltonian $\tilde{D}_{R}$ as 
\begin{equation}
\langle e_i | \tilde{D}_{R} | e_j \rangle \equiv 
\left\{ 
 \begin{array}{cc}
      \langle e_i | D_R | e_j \rangle =\delta_{ij} \text{Re} \lambda_i  & \text{for} \quad i \in A , \\
      0 &\text{for} \quad i \not\in A , \\ 
 \end{array}
\right. \label{DRtilder}
\end{equation}
and define $H_{\text{eff}}$ by 
\begin{equation}
H_{\text{eff}} \equiv P \tilde{D}_{R} P^{-1}.
\end{equation} 
Since $(\tilde{D}_{R})^{\dag} = \tilde{D}_{R}$, $H_{\text{eff}}$ is $Q$-hermitian, 
\begin{equation}
H_{\text{eff}} ^{\dag^Q} =H_{\text{eff}},
\end{equation} 
and satisfies $H_{\text{eff}} | \lambda_i \rangle = \text{Re} \lambda_i | \lambda_i \rangle$. 
Furthermore, we introduce $| \tilde\psi(t) \rangle \equiv \sum_{i \in A}  a_i(t) | \lambda_i \rangle $. 
Then $| \psi(t) \rangle$ is approximately estimated as 
\begin{eqnarray}
| \psi(t) \rangle 
&\simeq& e^{ \frac{1}{\hbar} B (t-t_0)} 
\sum_{i \in A}  a_i(t_0) e^{-\frac{i}{\hbar} {\text Re} \lambda_i (t-t_0)} | \lambda_i \rangle \nonumber\\
&=&e^{ \frac{1}{\hbar} B (t-t_0)}  e^{-\frac{i}{\hbar} H_{\text{eff}} (t-t_0)} | \tilde\psi(t_0) \rangle \nonumber\\
&=& | \tilde\psi(t) \rangle . \label{psiprimetket}
\end{eqnarray}
The factor $e^{ \frac{1}{\hbar} B (t-t_0)} $ in eq.(\ref{psiprimetket}) 
can be dropped out by normalization. 
Thus we have effectively obtained a $Q$-hermitian Hamiltonian $H_{\text{eff}}$ 
after a long time development though our theory is described 
by the non-hermitian Hamiltonian $H$ at first. 
Indeed the normalized state 
\begin{equation}
| \psi(t) \rangle_{N} 
\simeq \frac{1}{\sqrt{ \langle \tilde{\psi}(t) |_Q ~\tilde{\psi}(t) \rangle} } | \tilde{\psi}(t) \rangle \nonumber \equiv | \tilde{\psi}(t) \rangle_{N}
\end{equation} 
time-develops as $| \tilde{\psi}(t) \rangle_{N} =e^{-\frac{i}{\hbar} H_{\text{eff}} (t-t_0)} | \tilde{\psi}(t_0) \rangle_{N}$.
We see that the time dependence of the normalization factor 
has disappeared due to the $Q$-hermiticity of $H_{\text{eff}}$. 
Thus $| \tilde{\psi}(t) \rangle_{N}$, the normalized state 
by using the inner product $I_Q$, obeys the Schr\"{o}dinger equation  
\begin{equation}
i\hbar \frac{\partial}{ \partial t} | \tilde\psi(t) \rangle_{N} = H_{\text{eff}} | \tilde\psi(t) \rangle_{N}.
\end{equation}
On the other hand, the expectation value is given by 
\begin{equation}
\bar{\cal O}_Q(t) \simeq  {}_{N} \langle \tilde\psi(t) |_Q {\cal O} | \tilde\psi(t) 
\rangle_{N} 
= {}_{N} \langle \tilde\psi(t_0) |_Q \tilde{\cal O}_{QH}(t-t_0) | \tilde\psi(t_0) \rangle_{N},
\end{equation} 
where we have defined a time-dependent operator $\tilde{\cal O}_{QH}$ 
in the Heisenberg picture by 
\begin{equation}
\tilde{\cal O}_{QH}(t-t_0) \equiv    e^{ \frac{i}{\hbar} H_{\text{eff}} (t-t_0) } {\cal O}  e^{ -\frac{i}{\hbar} H_{\text{eff}}(t-t_0) }.
\end{equation}
We see that $\tilde{\cal O}_{QH}$ obeys the Heisenberg equation 
\begin{equation}
\frac{d}{dt} \tilde{\cal O}_{QH}(t-t_0) = \frac{i}{\hbar} [ H_{\text{eff}}, \tilde{\cal O}_{QH} (t-t_0)].
\end{equation}

\section{Summary and outlook}

In this paper we have proposed the replacement of hermitian operators of coordinate and momentum 
$\hat{q}$ and $\hat{p}$ and their eigenstates $\langle q |$ and $\langle p |$ 
with non-hermitian operators $\hat{q}_{new}$ and 
$\hat{p}_{new}$, and ${}_m \langle_{new}~ q |$ and ${}_m \langle_{new}~ p |$ with 
complex eigenvalues $q$ and $p$, 
so that we can express complex saddle points in terms of bras and kets.
We have formulated a complex action theory (CAT) 
such that mass and other coupling parameters are generically complex, 
while a coordinate $q$ and a momentum $p$ are fundamentally real, but can be complex 
at saddle points.

Indeed, in section~\ref{fundamental} 
to realize the philosophy of keeping the analyticity in dynamical variables of 
Feynman path integral (FPI), 
we have defined several new devices, that is to say, 
a modified set of complex conjugate $*_{ \{ \} }$, real and imaginary parts $\text{Re}_{\{ \}}$, $\text{Im}_{\{ \}}$, 
hermitian conjugate $\dag_{\{\}}$, and bras ${}_m\langle ~|$, ${}_{\{\}} \langle ~|$, 
where $\{\}$ denotes a set of parameters in which we keep the analyticity. 
We have also seen that the delta function can be used also for a complex parameter, 
when it satisfies such a condition as eq.(\ref{cond_of_q_for_delta}). 
In section 3 we have explicitly constructed the non-hermitian operators 
$\hat{q}_{new}$ and $\hat{p}_{new}$, and the eigenstates of their hermitian conjugates 
$|q \rangle_{new}$ and $|p \rangle_{new}$ with complex 
eigenvalues $q$ and $p$ by formally utilizing coherent states of harmonic oscillators. 
In appendix~\ref{app_cs} we have briefly reviewed a coherent state. 
Only in our formalism can we describe the CAT and 
a real action theory (RAT) with complex saddle points in the tunneling effect 
etc. in terms of bras and kets in the functional integral. 
In appendix~\ref{explicitcalc} we have explicitly studied various properties of 
$\hat{q}_{new}^\dag$, $\hat{p}_{new}^\dag$, $| q \rangle_{new}$ and $| p \rangle_{new}$. 
Especially we have seen that 
$\hat{q}_{new}^\dag$, $\hat{p}_{new}^\dag$, $|q \rangle_{new}$ and $|p \rangle_{new}$ behave in a similar way 
as $\hat{q}$, $\hat{p}$, $|q \rangle$ and $|p \rangle$, and 
we have the relations $\hat{p}_{new}^\dag | q \rangle_{new} = i \hbar \frac{\partial   | q \rangle_{new} }{\partial q}$ and 
$\hat{q}_{new}^\dag | p \rangle_{new} = \frac{\hbar}{i} \frac{\partial  | p \rangle_{new} }{\partial p} $ 
with complex $q$ and $p$ by insisting on $[\hat{q}_{new},  \hat{p}_{new} ] = i \hbar$. 
Furthermore, in appendix~\ref{theorem}, 
to make clear the relation between functions on the phase space describing some classical variables 
and the corresponding operators under quantization, 
providing the notions of ``$\epsilon$-analytical" functions, 
and ``expandable" and  ``$\epsilon$-expandable" operators, we have posed a theorem, 
which claims that if and only if some operator corresponding to an $\epsilon$-expandable 
function on the phase space, its matrix element in $q$-representation is an $\epsilon$-analytical function. 

In section~\ref{possible_mechanism}, as an application of 
the complex coordinate formalism 
which we have developed in the foregoing sections, we have extended 
our previous work \cite{Nagao:2010xu} to the complex coordinate formalism. 
We have studied a system defined by a non-hermitian diagonalizable bounded Hamiltonian $H$, 
and have shown that the framework presented in ref.\cite{Nagao:2010xu} for suppressing 
the effects of the anti-hermitian part of $H$ works also in the complex coordinate formalism. 
The framework is composed of two steps: 
As the first step we have introduced a proper inner product $I_Q$ such that 
the eigenstates of the Hamiltonian with different eigenvalues 
get orthogonal with regard to it, and also defined a hermiticity with regard to it, 
$Q$-hermiticity. With regard to $I_Q$ the Hamiltonian is normal. 
As the second step we have seen 
that the states with the highest imaginary part of 
the eigenvalues of $H$ get more favored to result than others after a long time development. 
Thus, the anti-$Q$-hermitian part of $H$ gets attenuated except for an unimportant constant, 
and we have effectively obtained a $Q$-hermitian Hamiltonian $H_{\text{eff}}$. 
This result suggests that we have {\it no reason to maintain that at the fundamental level the Hamiltonian should be hermitian}.

If $H$ is written in a local form, 
does the locality remain even after $H$ becomes the $Q$-hermitian Hamiltonian $H_{\text{eff}}$? 
It is not clear, but in ref.\cite{Nagao:2010xu} we have supposed 
that $H_{\text{eff}}$ has a local expression like 
$H_{\text{eff}} \simeq - \frac{\hbar^2}{2m_{\text{eff}}} 
\frac{\partial^2 }{\partial  q_{\text{eff}}^2 }
+ V_{\text{eff}}(q_{\text{eff}})$ 
with real $q_{\text{eff}}$, and have constructed a conserved probability current density. 
We can formally extend it to the complex $q_{\text{eff}}$ 
case.\footnote{A true probability interpretation stands for real $q_{\text{eff}}$.} 
Introducing two kinds of wave functions 
$\tilde{\psi}(q_{\text{eff}}) \equiv {}_m\langle_{new}~ q_{\text{eff}} | \tilde{\psi}(t) \rangle_N$ and 
$\tilde{\psi}_Q (q_{\text{eff}}) \equiv {}_m\langle_{new}~ q_{\text{eff}} |_Q~ \tilde{\psi}(t) \rangle_N$, 
we define a probability density by 
\begin{equation}
\rho_{\text{eff}}=\tilde{\psi}_Q(q_{\text{eff}})^{*_{q_{\text{eff}}}} \tilde{\psi}(q_{\text{eff}})
={}_N\langle \tilde{\psi}(t) |_Q~ q_{\text{eff}} \rangle_{new} 
~{}_m\langle_{new}~ q_{\text{eff}} | \tilde{\psi}(t) \rangle_N . \label{rho_eff}
\end{equation}
Then, since $\tilde{\psi}(q_{\text{eff}})$ and $\tilde{\psi}_Q(q_{\text{eff}})$ satisfy 
$i\hbar \frac{\partial}{\partial t} \tilde{\psi}(q_{\text{eff}}) = 
H_{\text{eff}} \tilde{\psi}(q_{\text{eff}})$ and 
$i\hbar \frac{\partial}{\partial t} \tilde{\psi}_Q(q_{\text{eff}}) = 
H_{\text{eff}}^{*_{q_{\text{eff}}}} \tilde{\psi}_Q(q_{\text{eff}})$ respectively, 
we obtain a continuity equation 
\begin{equation}
\frac{\partial \rho_{\text{eff}}}{\partial t} 
+ \frac{\partial}{\partial q_{\text{eff}}} j_{\text{eff}}(q_{\text{eff}},t) = 0, \label{cont_eq}
\end{equation}
where $j_{\text{eff}}(q_{\text{eff}},t)$ is a probability current density defined by 
\begin{equation}
j_{\text{eff}}(q_{\text{eff}},t)= \frac{i\hbar}{2 m_{\text{eff}}} 
\left( \frac{\partial}{\partial q_{\text{eff}}} \tilde{\psi}_Q(q_{\text{eff}})^{*_{q_{\text{eff}}}} 
\tilde{\psi}(q_{\text{eff}})       
- \tilde{\psi}_Q(q_{\text{eff}})^{*_{q_{\text{eff}}}} \frac{\partial}{\partial q_{\text{eff}}} 
\tilde{\psi}(q_{\text{eff}}) \right). \label{j_eff}
\end{equation} 
Thus if $H_{\text{eff}}$ has the local expression, we have the probability conservation 
$\frac{d}{dt} \int_C \rho_{\text{eff}} ~dq_{\text{eff}}=0$. 
Next we examine other possible candidates of a probability density and 
a probability current density. 
If we attempt to construct them only in terms of $\tilde{\psi}(q_{\text{eff}})$ as 
$\rho_{\text{eff}}=\tilde{\psi}(q_{\text{eff}})^{*_{q_{\text{eff}}}} \tilde{\psi}(q_{\text{eff}})$ and 
$j_{\text{eff}}(q_{\text{eff}},t) = \frac{i\hbar}{2 m_{\text{eff}}} 
\left( \frac{\partial}{\partial q_{\text{eff}}} \tilde{\psi}(q_{\text{eff}})^{*_{q_{\text{eff}}}} 
\tilde{\psi}(q_{\text{eff}}) 
- \tilde{\psi}(q_{\text{eff}})^{*_{q_{\text{eff}}}} \frac{\partial}{\partial q_{\text{eff}}} 
\tilde{\psi}(q_{\text{eff}}) \right)$, 
then this combination does not satisfy the continuity equation (\ref{cont_eq}).
Also, another pair written only in terms of $\tilde{\psi}_Q (q_{\text{eff}})$ as 
$\rho_{\text{eff}}=\tilde{\psi}_Q(q_{\text{eff}})^{*_{q_{\text{eff}}}} \tilde{\psi}_Q(q_{\text{eff}})$ and 
$j_{\text{eff}}(q_{\text{eff}},t) = \frac{i\hbar}{2 m_{\text{eff}}} 
\left( \frac{\partial}{\partial q_{\text{eff}}} \tilde{\psi}_Q (q_{\text{eff}})^{*_{q_{\text{eff}}}} 
\tilde{\psi}_Q(q_{\text{eff}})       
- \tilde{\psi}_Q(q_{\text{eff}})^{*_{q_{\text{eff}}}} \frac{\partial}{\partial q_{\text{eff}}} 
\tilde{\psi}_Q(q_{\text{eff}}) \right)$ 
does not satisfy the continuity equation (\ref{cont_eq}). 
Only the combination of eqs.(\ref{rho_eff})(\ref{j_eff}) satisfy eq.(\ref{cont_eq}). 
We also note that eqs.(\ref{rho_eff})(\ref{j_eff}) are not defined locally 
due to the existence of $Q$. 
The detail study of $Q$ is an open problem. 


Now we have the philosophy of keeping the analyticity in FPI parameters, 
new devices to realize it, 
non-hermitian operators $\hat{q}_{new}$ and $\hat{p}_{new}$, and their eigenstates ${}_m\langle_{new}~  q | $ and ${}_m\langle_{new}~ p | $ with complex eigenvalues $q$ and $p$, 
so we can go ahead to study the CAT further in detail. 
We expect that the philosophy and the new devices which we have introduced in this paper 
would be useful for studying the properties and dynamics of the CAT 
and even the RAT with complex saddle points in the tunneling effect 
etc. in terms of bras and kets in the functional integral. 
As a next step, what should we study to develop the CAT?
First, we note that in this paper we have not explicitly studied the classical behavior. 
We have assumed that the correspondence principle 
between a quantum regime and a classical one holds in our system. 
At the point where the imaginary part of the action $S_I$ is minimized,
we have $\delta S_I \simeq 0$, so that in the region around it
$S_I$ is constant practically.
Thus we see little effect of $S_I$ there.
This is consistent with our observation that the anti-hermitian part
of the Hamiltonian is suppressed after a long time. 
It is desirable to study it somehow. 
Second, a conserved probability current density, which we have constructed 
under the assumption that $H_{\text{eff}}$ is given in a local form with some complex coordinate $q_{\text{eff}}$, 
is not defined locally due to the existence of $Q$. 
It is important to study $Q$ in detail. 
Also, in ref.\cite{Nagao:2010xu} we have pointed out a possible misestimation of an early state 
by extrapolating back in time with the hermitian Hamiltonian $H_{\text{eff}}$. 
Though we have not discussed it in this paper, it would be interesting to investigate it in detail. 
Furthermore, in this paper we have not considered a future-included theory, that is to say, 
a theory including not only a past time but also a future time as an integration 
interval of time. 
In ref.\cite{Nielsen:2007mj} a kind of wave function of universe 
including the information of future is introduced. 
It is intriguing to study such a future-included theory in the complex coordinate formalism. 
We will study them and report the progress in the future.



\section*{Acknowledgements}

This work was partially funded by Danish Natural Science Research Council (FNU, Denmark), 
and the work of one of the authors (K.N.) was supported in part by 
Grant-in-Aid for Scientific Research (Nos.18740127 and 21740157) 
from the Ministry of Education, Culture, Sports, Science and Technology (MEXT, Japan). 
K.N. would like to thank all the members and visitors of NBI for their kind hospitality.  
In addition, the authors are very grateful to the referees of Progress of Theoretical Physics for 
inspiring them to improve the manuscript into this revised version.

\appendix

\section{Coherent state}\label{app_cs}

We briefly summarize the $q$ and $p$-representations of a coherent state. 
The coherent state parametrized with a complex parameter $\lambda$ 
is defined up to a normalization factor by 
\begin{equation}
| \lambda \rangle_{coh} \equiv  e^{\lambda a^\dag} | 0 \rangle 
= \sum_{n=0}^{\infty} \frac{\lambda^n}{\sqrt{n!}} | n \rangle , \label{lambda_coh_def}
\end{equation}
and satisfies the relation 
\begin{equation}
a | \lambda \rangle_{coh} = \lambda | \lambda \rangle_{coh} . \label{a_lambda_ket=lambda_lambda_ket}
\end{equation}
In eqs.(\ref{lambda_coh_def})(\ref{a_lambda_ket=lambda_lambda_ket}) $a$ and $a^\dag$ are 
annihilation and creation operators defined by 
\begin{eqnarray}
&&a = \sqrt{ \frac{m\omega}{2\hbar}}  \left( \hat{q} + i \frac{ \hat{p}}{m \omega}  \right) , \label{annihilation} \\
&&a^\dag = \sqrt{ \frac{m\omega}{2\hbar}}\left( \hat{q} - i \frac{ \hat{p}}{m \omega}  \right) \label{creation}.
\end{eqnarray}
The eigenstates of $\hat{q}$ and $\hat{p}$ are $| q \rangle$ and $| p \rangle$ respectively, 
and they obey 
\begin{eqnarray}
&&\hat{q} | q \rangle = q| q \rangle , \label{qhatqket=qqket} \\
&&\hat{p} | p \rangle = p | p \rangle ,    \label{phatpket=ppket} \\
&&\hat{q} | p \rangle = \frac{\hbar}{i} \frac{\partial}{\partial p}  | p \rangle , \label{qhatpket=ihbardeldelppket} \\ 
&&\hat{p} | q \rangle = i \hbar \frac{\partial}{\partial q}  | q \rangle ,  \label{phatqket=ihbardeldelqqket} \\
&&\langle q | p \rangle = \frac{1}{\sqrt{2\pi \hbar}} e^{\frac{i}{\hbar} pq} , \label{qbrapket=eihbarpq_real} \\ 
&&[ \hat{q},  \hat{p} ] = i \hbar. \label{commu_rel_qhatphat}
\end{eqnarray}
The $q$ and $p$-representations of the coherent state are given by 
\begin{eqnarray}
\langle q | \lambda \rangle_{coh} 
&=& \left( \frac{m\omega}{\pi \hbar} \right)^{\frac{1}{4}} 
e^{ \frac{1}{2} \lambda^2}
\exp\left[  -\frac{m\omega}{2\hbar} \left( q - \lambda \sqrt{ \frac{2\hbar}{m \omega} }  \right)^2 \right] \nonumber \\
&=&
\left( \frac{m\omega}{\pi \hbar} \right)^{\frac{1}{4}} 
e^{ \frac{m\omega}{4 \hbar} q_0^2} 
e^{  \frac{p_0^2}{ 4 \hbar m \omega } }  
e^{ - i \frac{p_0 q_0}{2 \hbar} }
\exp\left[  -\frac{m\omega}{2\hbar} \left( q - q_0 \right)^2 \right] 
\exp\left[  \frac{i}{\hbar} p_0 q \right] , \label{q_rep_lambda_state} \\
\langle p | \lambda \rangle_{coh} 
&=& \left( \frac{1}{\pi \hbar m\omega} \right)^{\frac{1}{4}} 
e^{ -\frac{1}{2} \lambda^2}
\exp\left[ -\frac{1}{2\hbar m\omega} \left( p + i \lambda \sqrt{ 2\hbar m \omega } \right)^2 \right]  \nonumber \\
&=&
\left( \frac{1}{\pi \hbar m\omega} \right)^{\frac{1}{4}} 
e^{ \frac{m\omega}{4 \hbar} q_0^2} 
e^{  \frac{p_0^2}{ 4 \hbar m \omega } }  
e^{ i \frac{p_0 q_0}{2 \hbar} }
\exp\left[  -\frac{1}{2\hbar m\omega } \left( p - p_0 \right)^2 \right] 
\exp\left[  -\frac{i}{\hbar} q_0 p \right] , \label{p_rep_lambda_state} \nonumber \\
\end{eqnarray}
where we have introduced $q_0$ and $p_0$ as 
\begin{eqnarray}
&&q_0 = \sqrt{ \frac{2 \hbar}{m \omega}}  ~\text{Re} \lambda  , \label{q_0} \\
&&p_0 = \sqrt{ 2\hbar m \omega }  ~\text{Im} \lambda  . \label{p_0}
\end{eqnarray}
Eqs.(\ref{q_rep_lambda_state})(\ref{p_rep_lambda_state}), up to the normalization, show that in the phase space $(q, p)$ 
the coherent state is expressed as a wave packet 
located around $(q_0, p_0)$ with the widths $\Delta q= \sqrt{ \frac{\hbar}{m \omega} }$ 
and $\Delta p = \sqrt{ \hbar m \omega }$ in the $q$ and $p$ directions respectively.

\section{Explicit studies on the properties of $\hat{q}_{new}^\dag$, $\hat{p}_{new}^\dag$, $|q \rangle_{new}$ 
and $|p \rangle_{new}$}\label{explicitcalc}

In this section, as a supplement to section~\ref{construction}, 
we study various properties of $\hat{q}_{new}$, $\hat{p}_{new}$, $|q \rangle_{new}$ and $|p \rangle_{new}$, 
which are defined in eqs.(\ref{def_qhat_new})(\ref{def_phat_new})(\ref{def_qketnew})(\ref{def_pketnew}) respectively.

\subsection{Completeness relations for $|q \rangle_{new}$ and $|p \rangle_{new}$}\label{completeness}

The integral $\int_C dq | q \rangle_{new} ~{}_m\langle_{new} ~q |$ is calculated as 
\begin{eqnarray}
&&\int_C dq | q \rangle_{new} ~{}_m \langle_{new} ~q | \nonumber \\
&=& \frac{m\omega}{2 \pi \hbar} \sqrt{1 - \frac{m' \omega'}{m\omega} }
\int_{real} dq' dq'' | q' \rangle \langle q'' | \nonumber \\
&&\times \int_C dq 
\exp\left[ - \frac{m\omega}{\hbar} \left( 1 - \frac{m' \omega'}{m\omega} \right) 
\left( q- \frac{q' + q''}{2 \sqrt{1 - \frac{m' \omega'}{m\omega} }} \right)^2 \right] 
\exp\left[ -  \frac{m\omega}{4\hbar} (q' - q'')^2 \right] \nonumber \\
&=&
\int_{real} dq' dq''| q' \rangle \langle q'' | \delta^{\epsilon_3}_c (q'-q'') \nonumber \\
&=& 1 , \label{completion_complexq_ket}
\end{eqnarray}
where we have introduced $\epsilon_3 = \frac{\hbar}{m\omega}$. 
Similarly, the integral $\int_C dp | p \rangle_{new} ~{}_m\langle_{new} ~p |$ is estimated as 
\begin{equation}
\int_C dp | p \rangle_{new} ~{}_m \langle_{new} ~p | = 1 . \label{completion_complexp_ket}
\end{equation}
Thus we have the completeness relations for $|q \rangle_{new}$ and $|p \rangle_{new}$.

\subsection{Properties of $\hat{q}_{new}^\dag$ and $\hat{p}_{new}^\dag$}\label{qpnewhatreal}

Since $\hat{q}_{new}^\dag$ and $\hat{p}_{new}^\dag$ obey the following relations, 
\begin{eqnarray}
&&\hat{q}_{new}^\dag | q' \rangle \simeq  \hat{q}  | q' \rangle  \quad 
\text{for large $m \omega$ and real $q'$} , \\
&&\hat{q}_{new}^\dag | p' \rangle 
\simeq \hat{q} | p' \rangle  \quad 
\text{for large $m \omega$ and real $p'$} , \\
&&
\hat{p}_{new}^\dag | q' \rangle 
\simeq  \hat{p}  | q' \rangle  \quad 
\text{for small $m' \omega'$ and real $q'$} , \\
&&\hat{p}_{new}^\dag | p' \rangle \simeq  \hat{p}  | p' \rangle  \quad 
\text{for small $m' \omega'$ and real $p'$} , 
\end{eqnarray}
we see that $\hat{q}_{new}^\dag$ and $\hat{p}_{new}^\dag$ behave like 
$\hat{q}$ and $\hat{p}$ respectively for  $| q' \rangle$ with real $q'$ or $| p' \rangle$ with real $p'$. 
I.e. we have 
\begin{eqnarray}
&&\hat{q}_{new}^\dag \simeq \hat{q}  \quad \text{for large $m \omega$,   and for $| q' \rangle$ with real $q'$ or $| p' \rangle$ with real $p'$} , \\
&&\hat{p}_{new}^\dag \simeq \hat{p}  \quad \text{for small $m' \omega'$, and for $| q' \rangle$ with real $q'$ or $| p' \rangle$ with real $p'$} .
\end{eqnarray}

\subsection{Expressions of $\langle q' | q \rangle_{new}$ and $ \langle p' | p \rangle_{new} $ 
with real $q'$ and $p'$}\label{q'ketp'ket}

In eqs.(\ref{braq'qketnew2})(\ref{brap'pketnew2}) we have given the explicit expressions of 
$\langle q' | q \rangle_{new}$ and $\langle p' | p \rangle_{new}$, 
but they are not written in a manner to keep the analyticity in $q$ and $p$, 
so we give other expressions. 
We rewrite $\langle q' | q \rangle_{new}$ for real $q'$ as 
\begin{eqnarray}
\langle q' | q \rangle_{new} 
&=& 
\left\{ \frac{m\omega}{4 \pi \hbar} \left( 1 - \frac{m'\omega'}{m\omega} \right) \right\}^{\frac{1}{4}} 
e^{- \frac{m\omega}{4\hbar} \left( 1 - \frac{m'\omega'}{m\omega} \right) {q}^2 }
\langle q' | \sqrt{ \frac{m\omega}{2\hbar} 
\left( 1 - \frac{m'\omega'}{m\omega} \right) } q \rangle_{coh}  \nonumber \\
&=& 
\left( 1 - \frac{m'\omega'}{m\omega} \right)^{\frac{1}{4}} 
\delta_c^{\epsilon_2} \left( q' - \sqrt{ 1 - \frac{m'\omega'}{m\omega}} q \right) \nonumber \\
&\simeq& \delta_c^{\epsilon_2} \left( q' - q \right) \quad \text{for large $m \omega$} , 
\label{braq'qketnew}
\end{eqnarray}
where in the second equality we have used eq.(\ref{q_rep_lambda_state}) and introduced 
\begin{equation}
\epsilon_2 = \frac{\hbar}{2m \omega} .   \label{epsilon_2}
\end{equation}
The tamed delta function converges for $q'$ and $q$ satisfying 
$L\left( q' - \sqrt{ 1 - \frac{m'\omega'}{m\omega}} q \right)>0$. 
Similarly $\langle p' | p \rangle_{new}$ with real $p'$ is estimated as 
\begin{eqnarray}
\langle p' | p \rangle_{new} 
&=& 
\left( 1 - \frac{m'\omega'}{m\omega} \right)^{\frac{1}{4}} 
\delta_c^{\epsilon'_2} \left( p' - \sqrt{ 1 - \frac{m'\omega'}{m\omega}} p \right) \nonumber \\
&\simeq& \delta_c^{\epsilon'_2} \left( p' - p \right) \quad \text{for small $m' \omega'$} , 
\label{brap'pketnew}
\end{eqnarray}
where in the first equality we have used eq.(\ref{p_rep_lambda_state}) and introduced 
\begin{equation}
\epsilon'_2 = \frac{\hbar m' \omega'}{2} .  \label{epsilon'_2}
\end{equation}
The tamed delta function converges for $p'$ and $p$ satisfying 
$L\left( p' - \sqrt{ 1 - \frac{m'\omega'}{m\omega}} p \right)>0$.

In addition we write $| q \rangle_{new}$ and $| p \rangle_{new}$ 
as superpositions of $| q' \rangle$ and $| p' \rangle$ with real $q'$ and $p'$ as follows, 
\begin{eqnarray}
&&| q \rangle_{new} = \int_{\text{real axis}} \langle q' | q \rangle_{new}  | q' \rangle dq' ,  
\label{qketnew_exp_qprimeket} \\
&&
| p \rangle_{new} = \int_{\text{real axis}} \langle p' | p \rangle_{new}  | p' \rangle dp' . 
\label{pketnew_exp_pprimeket}
\end{eqnarray}
Then using eqs.(\ref{braq'qketnew})(\ref{brap'pketnew}) we see that the $| q' \rangle_{new}$ 
with real $q'$ and $| p' \rangle_{new}$ with real $p'$ become $| q' \rangle$ and $| p' \rangle$ respectively, 
\begin{eqnarray}
&& | q' \rangle_{new} \simeq | q' \rangle   \quad \text{for large $m \omega$ and real $q'$} , 
\label{qketnew_exp_qprimeketreal} \\
&& | p' \rangle_{new} \simeq | p' \rangle  \quad \text{for small $m' \omega'$ and real $p'$} . 
\label{pketnew_exp_pprimeketreal}
\end{eqnarray}

\subsection{Analyses of $\hat{p}_{new}^\dag | q \rangle_{new}$ and $\hat{q}_{new}^\dag | p \rangle_{new}$} \label{phatqket_qhatpket}

The ket $\hat{p}_{new}^\dag| q \rangle_{new}$ is calculated as follows, 
\begin{eqnarray}
\hat{p}_{new}^\dag| q \rangle_{new} 
&=&
\frac{1}{ \sqrt{1 - \frac{m' \omega'}{m\omega} } } 
\int_{\text{real axis}} dq' 
\left\{     
\frac{\hbar}{i} \left( \frac{\partial}{\partial q'} \langle q' | q \rangle_{new} \right) 
| q' \rangle
-i m' \omega' q' \langle q' | q \rangle_{new} | q' \rangle
\right\} 
\nonumber \\
&=&
-i q m \omega \int_{\text{real axis}} dq' 
\langle q' | q \rangle_{new} | q' \rangle
+ 
i m \omega \sqrt{1 - \frac{m' \omega'}{m\omega} } 
\int_{\text{real axis}} dq' 
q' \langle q' | q \rangle_{new} | q' \rangle \nonumber \\
&\simeq&i \hbar \frac{\partial}{\partial q} | q \rangle_{new} \quad \text{for small $m' \omega'$} , 
\label{phatnewqketnew}
\end{eqnarray}
where in the first equality we have used eqs.(\ref{qhatqket=qqket})(\ref{phatqket=ihbardeldelqqket}) 
for real $q$ and $p$, and 
performed a partial integral.  
In the second and third equalities we have utilized 
\begin{eqnarray}
&&\frac{\partial}{\partial q'} \langle q' | q \rangle_{new}
= 
-\frac{m\omega}{\hbar} \left( q' - \sqrt{1 - \frac{m' \omega'}{m\omega} } q \right)
\langle q' | q \rangle_{new} , \\
&&q' \langle q' | q \rangle_{new} 
= \sqrt{1 - \frac{m' \omega'}{m\omega} } q \langle q' | q \rangle_{new} 
+\frac{\hbar}{m\omega} \frac{1}{ \sqrt{1 - \frac{m' \omega'}{m\omega} }} 
\frac{\partial}{\partial q} \langle q' | q \rangle_{new} , 
\end{eqnarray}
respectively. 
Also, $\hat{q}_{new}^\dag| p \rangle_{new}$ is similarly calculated as 
\begin{equation}
\hat{q}_{new}^\dag| p \rangle_{new} 
\simeq \frac{\hbar}{i} \frac{\partial}{\partial p} | p \rangle_{new} \quad 
\text{for large $m \omega$} . \label{qhatnewpketnew}
\end{equation}

\subsection{Analyses of a basis function of the Fourier transformation}\label{basisfourier}

A basis function of the Fourier transformation is calculated as  
\begin{eqnarray}
&&{}_m\langle_{new} ~q | p \rangle_{new} \nonumber \\
&=&
\int_{C} dq' \int_{C} dp' 
\sqrt{1 - \frac{m' \omega'}{m\omega} }
\delta_c^{\epsilon_2} \left(q'- \sqrt{1 - \frac{m' \omega'}{m\omega} } q \right) 
\delta_c^{\epsilon'_2} \left(p'- \sqrt{1 - \frac{m' \omega'}{m\omega} } p \right) 
\frac{1}{\sqrt{2 \pi \hbar}} e^{\frac{i}{\hbar}p' q'} \nonumber \\
&\simeq&
\frac{1}{\sqrt{2 \pi \hbar}} e^{\frac{i}{\hbar}p q}  \quad \text{for large $m\omega$ and small $m'\omega'$} , 
\label{basis_fourier_transf}
\end{eqnarray}
where in the first equality $\epsilon_2$ and $\epsilon'_2$ are given in eqs.(\ref{epsilon_2})(\ref{epsilon'_2}) respectively. 
Accordingly ${}_m\langle_{new} ~q | p \rangle_{new}$ has a similar expression to eq.(\ref{qbrapket=eihbarpq_real}). 
The Fourier transformation is formally defined in the CAT with this basis function. 
In eqs.(\ref{brap'qketnew2})(\ref{braq'pketnew2}) 
$\langle p' | q \rangle_{new}$ and $\langle q' | p \rangle_{new}$ are expressed for real $q'$ and $p'$, 
but their analyticities are not kept in $q$ and $p$. 
We rewrite $\langle p' | q \rangle_{new}$ keeping the analyticity in $q$ as follows, 
\begin{eqnarray}
\langle p' | q \rangle_{new} 
&=& 
\frac{1}{ \sqrt{ 2\pi \hbar} }
\left( 1 - \frac{m'\omega'}{m\omega} \right)^{\frac{1}{4}} 
\exp\left[  -\frac{{p'}^2}{2\hbar m\omega} \right]
\exp\left[  - \frac{i}{\hbar} \sqrt{ 1 - \frac{m'\omega'}{m\omega}} p' q \right]  \nonumber \\
&\simeq& 
\frac{1}{ \sqrt{ 2\pi \hbar} } \exp\left[  - \frac{i}{\hbar} p' q \right] 
\quad \text{for large $m \omega$} . \label{brap'qketnew_app} 
\end{eqnarray}
Similarly, 
$\langle q' | p \rangle_{new}$ and ${}_m \langle_{new} ~q | p' \rangle$ are expressed as 
\begin{eqnarray}
&&\langle q' | p \rangle_{new} 
\simeq
\frac{1}{ \sqrt{ 2\pi \hbar} } \exp\left[  \frac{i}{\hbar} q' p \right] 
\quad \text{for small $m' \omega'$} , \label{braq'pketnew_app} \\
&&
{}_m \langle_{new} ~q | p' \rangle = (   \langle p' | q \rangle_{new}   )^{*_q} 
\simeq \frac{1}{ \sqrt{ 2\pi \hbar} } \exp\left[  \frac{i}{\hbar} p' q \right] \quad 
\text{for large $m \omega$} . \label{mbranewqp'ket_app}
\end{eqnarray}
Therefore $\langle q' | p \rangle_{new}$ and ${}_m \langle_{new} ~q | p' \rangle$ 
have similar expressions to eq.(\ref{qbrapket=eihbarpq_real}).

\section{The relation between functions and operators } \label{theorem}

In this section introducing ``$\epsilon$-analytical" functions, 
and ``expandable" and  ``$\epsilon$-expandable" operators, 
we pose a theorem on the relation between them. 
To proceed with this we first introduce an ``expandable" operator.

\subsection{An ``expandable" operator }

If a function ${\cal O} (q, p)$ is analytical in $q$ and $p$, i.e. it can be Taylor-expanded as 
${\cal O} (q,p) = \sum_{m,n} a_{mn} q^m p^n$, 
then we call the operator 
\begin{equation}
{\cal O} ( \hat{q}_{new}, \hat{p}_{new})= \sum_{m,n} a_{mn} (\hat{q}_{new})^m (\hat{p}_{new})^n \label{expandable_operator}
\end{equation}
``expandable" in $\hat{q}_{new}$ and $\hat{p}_{new}$. 
This operator is obtained by the replacement of $q$ and $p$ with $\hat{q}_{new}$ and $\hat{p}_{new}$, respectively. 
Here we have taken the ordering to be that $(\hat{q}_{new})^m$ was put to the left of $(\hat{p}_{new})^n$, but 
if we choose some other ordering convention, we can correct 
the coefficient $a_{mn}$ to other one $\tilde{a}_{mn}$, 
and still get an expanded expression like eq.(\ref{expandable_operator}). 
Indeed we can make a set of ``expandable" operators by considering all possible ordering of 
$\hat{q}_{new}$ and $\hat{p}_{new}$, but the set does not depend on the ordering convention.

\subsection{An ``$\epsilon$-analytical" function and an ``$\epsilon$-expandable" operator }

We also introduce a notion of an ``$\epsilon$-analytical" function.
For example, the delta function $\delta_c(q)$ is not an analytical function, 
but the tamed delta function $\delta_c^\epsilon(q)$ with finite $\epsilon$ is an analytical function. 
Thus we call $\delta_c(q)$ ``$\epsilon$-analytical" in the sense that 
it is analytical if we keep $\epsilon$ finite. 
As another example we consider a function $\frac{1}{q}$. 
This is not an analytical function, but can be expressed as 
$\frac{1}{q} = \lim_{\epsilon \rightarrow 0} \frac{1}{q + i \epsilon }$ 
by introducing a cutoff $\epsilon$. 
Thus, since $\frac{1}{q + i \epsilon }$ is analytical on the real axis of $q$, 
we can say that $\frac{1}{q}$ is $\epsilon$-analytical. 
Furthermore, if we make an operator ${\cal O} ( \hat{q}_{new}, \hat{p}_{new})$ 
by the replacement of $q$ and $p$ 
with $\hat{q}_{new}$ and $\hat{p}_{new}$ respectively in an $\epsilon$-analytical function, 
we call it an ``$\epsilon$-expandable" operator.

\subsection{A theorem on the relation between an $\epsilon$-analytical function and an $\epsilon$-expandable operator}

We have defined expandable and  $\epsilon$-expandable operators ${\cal{O}}( \hat{q}_{new} , \hat{p}_{new} )$ 
from analytical and $\epsilon$-analytical functions ${\cal O} (q,p)$. 
Then, it seems natural to wonder what are the corresponding properties of 
${}_m \langle_{new}~ q' | {\cal{O}}( \hat{q}_{new} , \hat{p}_{new} ) | q'' \rangle_{new}$. 
Is it in some sense analytical? 
To make it clear to some extent, we pose the following theorem. 

\vspace{0.5cm}

\noindent
{\bf Theorem}

\noindent
If and only if $\hat{\cal{O}}( \hat{q}_{new} , \hat{p}_{new})$ is an $\epsilon$-expandable operator, \\
${\cal{O}}( q' , q'') \equiv {}_m \langle_{new}~ q' | \hat{\cal{O}}( \hat{q}_{new} , \hat{p}_{new}) | q'' \rangle_{new}$ 
is an $\epsilon$-analytical function in $q'$ and $q''$.

\vspace{0.5cm}

We prove the theorem. If $\hat{\cal{O}} ( \hat{q}_{new} , \hat{p}_{new})$ is an $\epsilon$-expandable operator, 
we can express it in an expanded form like 
\begin{equation}
\hat{\cal O} ( \hat{q}_{new}, \hat{p}_{new})= \sum_{m,n} b_{mn} (\hat{q}_{new} )^m (\hat{p} _{new})^n . 
\end{equation}
Then for any finite value of $\epsilon$ we easily see 
\begin{eqnarray}
{\cal O}( q' , q'')
&=&
{}_m \langle_{new}~ q' | \hat{\cal O}( \hat{q}_{new} , \hat{p}_{new}) | q'' \rangle_{new} \nonumber \\
&=& 
\sum_{m,n} b_{mn} \left(  \frac{\hbar}{i} \frac{\partial}{\partial q'} \right)^n 
\left\{ (q')^m  \delta_c(q'-q'') \right\} . \label{Oqq_proof}
\end{eqnarray}
This is $\epsilon$-analytical in $q'$ and $q''$, 
so we have proven one direction of the theorem.

To prove the opposite direction of the theorem 
we attempt to see how $\hat{\cal{O}} ( \hat{q}_{new} , \hat{p}_{new})$ is expressed in terms of ${\cal{O}}( q' , q'')$. 
For this purpose we rewrite ${\cal{O}}( q' , q'')$ as follows, 
\begin{eqnarray}
{\cal{O}}( q' , q'') 
&=& \int {\cal{O}}( q', q'+a)  \delta_c(q' + a - q'' ) da \nonumber \\
&=& {}_m \langle_{new}~ q' | \int {\cal{O}}( q', q'+a) \exp\left( \frac{i}{\hbar} a \hat{p}_{new} \right) da | q'' \rangle_{new} 
\nonumber \\
&=& {}_m \langle_{new}~ q' | \int  
{\cal{O}}( \hat{q}_{new} , \hat{q}_{new} +a )  \exp\left( \frac{i}{\hbar} a \hat{p}_{new} \right) da | q'' \rangle_{new} , 
\label{Oqq_op_proof}
\end{eqnarray}
where in the second equality we have used the following relation, 
\begin{eqnarray}
{}_m \langle_{new}~ q' | \exp\left( \frac{i}{\hbar} a \hat{p}_{new}  \right)  | q'' \rangle_{new} 
&=& \sum_n \frac{ a^n}{n!} \frac{\partial^n }{ \partial (q')^n } \delta_c(q'  -q'') \nonumber \\
&=& \delta_c(q' + a -q'') .
\end{eqnarray}
From the last expression of eq.(\ref{Oqq_op_proof}), we can identify the corresponding operator 
$\hat{\cal{O}} ( \hat{q}_{new} , \hat{p}_{new})$ as follows, 
\begin{equation}
\hat{\cal{O}} ( \hat{q}_{new} , \hat{p}_{new}) =\int  
{\cal{O}}( \hat{q}_{new} , \hat{q}_{new} +a )  \exp\left( \frac{i}{\hbar} a \hat{p}_{new} \right)   da  .
\end{equation} 
If ${\cal{O}}( q' , q'')$ is an $\epsilon$-analytical function, 
${\cal{O}}( \hat{q}_{new} , \hat{q}_{new} +a )$  is 
an $\epsilon$-expandable operator, so $\hat{\cal{O}} ( \hat{q}_{new} , \hat{p}_{new})$ 
is an $\epsilon$-expandable operator. 
Thus we have proven the opposite direction of the theorem.


\end{document}